\documentclass{article}

\usepackage{arxiv}

\usepackage[utf8]{inputenc} % allow utf-8 input
\usepackage[T1]{fontenc}    % use 8-bit T1 fonts
\usepackage{url}            % simple URL typesetting
\usepackage{booktabs}       % professional-quality tables
\usepackage{amsfonts}       % blackboard math symbols
\usepackage{nicefrac}       % compact symbols for 1/2, etc.
\usepackage{microtype}      % microtypography

\usepackage{mathtools}

\usepackage{graphicx}

\usepackage{caption}

\usepackage{subcaption}

\usepackage{amssymb}
\usepackage{xcolor}

\usepackage{algorithm} 
\usepackage{algpseudocode}

\usepackage{hyperref}
\usepackage{doi}

\title{Lightweight Cryptanalysis of IoT Encryption Algorithms : Is Quota Sampling the Answer?}

\author{Jonathan Cook*,
        Sabih ur Rehman*
        and M. Arif Khan*\\ % <-this % stops a space\\
        {*School of Computing, Mathematics and Engineering, Charles Sturt University, Australia}}
\begin{document}
\maketitle

\begin{abstract}
	Rapid growth in the number of small sensor devices known as the Internet of Things (IoT) has seen the development of lightweight encryption algorithms. Two well-known lightweight algorithms are SIMON and SIMECK which have been specifically designed for use on resource-constrained IoT devices. These lightweight encryption algorithms are based on the efficient Feistel block structure which is known to exhibit vulnerabilities to differential cryptanalysis. Consequently, it is necessary to test these algorithms for resilience against such attacks. While existing state-of-the-art research has demonstrated novel heuristic methods of differential cryptanalysis that improve time efficiency on previous techniques, the large state sizes of these encryption algorithms inhibit cryptanalysis time efficiency. In this paper, we introduce Versatile Investigative Sampling Technique for Advanced Cryptanalysis (VISTA-CRYPT) - a time-efficient enhancement of differential cryptanalysis of lightweight encryption algorithms. The proposed technique introduces a simple framework of quota sampling that produces state-of-the-art results with time reductions of up to $76\%$ over existing techniques. Further, we present a preliminary graph-based analysis of the output differentials for the identification of relationships within the data and future research opportunities to further enhance the performance of differential cryptanalysis. The code designed for this work and associated datasets will be available at https://github.com/johncook1979/simon-cryptanalysis.
\end{abstract}

% keywords can be removed

\section{Introduction}\label{Sec:Introduction}

 As the world grows increasingly reliant on the Internet of Things (IoT), it is becoming critical to ensure that the data shared via IoT devices remains secure from adversarial attacks. A key component required to ensure data integrity is the integration of a robust cypher on IoT devices. Low-powered IoT devices, particularly those operating on limited battery cell power for prolonged periods of time, have traditionally had insufficient processing power necessary to run complex cyphers \cite{cook2023security}. Recognising the threat to the data collected and conveyed by IoT devices, the United States National Security Agency (NSA) developed two lightweight cyphers in 2013 to meet the security requirements of IoT devices \cite{beaulieu2013simon}. Based on the Rijndael Advanced Encryption Standard (AES), these two cyphers, SIMON and SPECK \cite{beaulieu2013simon}, sought to address the security constraints of both the processing and software limitations of IoT devices. Two years later in 2015, researchers at the University of Waterloo in Canada proposed a new lightweight cypher that comprised properties of both the SIMON and SPECK algorithms, creating a more efficient hardware cypher known as SIMECK \cite{yang2015simeck}. The new lightweight block cyphers use a simple operation known as bitwise AND rotation, also known as a circular shift \cite{stallings2017principles}. However, the development of cyphers to protect data does not guarantee that the cypher is impervious to adversarial attacks, and consequently, a necessary component of security research is the development and analysis of cryptanalysis techniques designed to find and exploit deficiencies and limitations within the cyphers \cite{edgar2017science}. Consequently, researchers have focused their attention on identifying methods to exploit these lightweight block cyphers to understand their limitations and weaknesses and found that for cryptanalysis to be considered effective it must be efficient \cite{de2006introduction}. According to the authors of \cite{easttom2021cryptanalysis}, three primary resources are consumed during cryptanalysis, which are the time taken to perform the cryptanalysis process, the amount of storage consumed as memory and the quantity of data consumed as cyphertext and plain text.

 The existing state-of-the-art technique proposed by the authors of \cite{dwivedi2023security} uses a heuristics Nested Monte-Carlo Search (NMCS) algorithm for differential cryptanalysis of the SIMON and SIMEK cyphers. Their proposed solution offers improved performance over existing cryptanalysis techniques, with times for smaller block sizes measured in minutes, not hours or days. The process of a NMCS is to conduct nested searches of a tree-like structure to identify the lowest hamming weight in the fastest possible time. Each node on a branch is assigned a weight and then added to the previous branch to determine the overall weight for that nested path. Each path is then compared by weight to determine the path with the lowest cumulative weight. By investigating smaller paths and then comparing the results, significant gains are made in the speed of the search. The solution proposed in their work provides an improvement on existing models for the investigation of differential paths within the SIMON and SIMECK cyphers.

\begin{figure*}[t]
    \centering
    \includegraphics[width=0.95\textwidth]{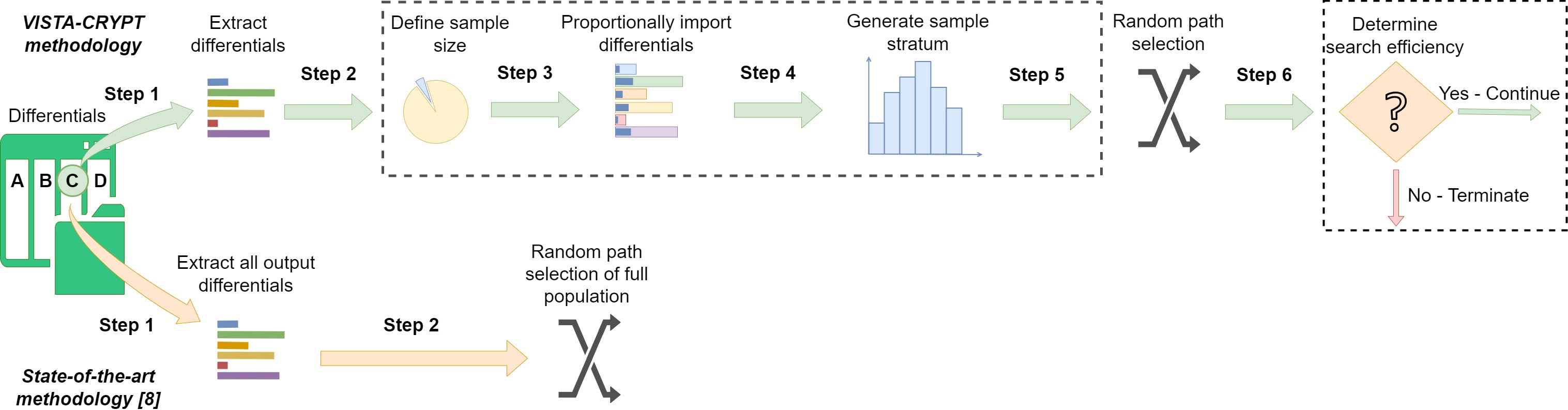}
    \caption{Our methodology: $1$) \textit{Extract output differentials:} Extract output differentials (C) from lists that also contain left input (A), right input (B) and weight (D). $2$) \textit{Define proportional sample size:} Define the proportion of differentials to use in the sample. $3$) \textit{Distribution extraction:} Extract the differentials with a minimum of one of each type to the sample. $4$) \textit{Sample generation:} A sample based on a quota is generated. $5$) \textit{Random path selection:} A random path from the sample is chosen. $6$) \textit{Decision on search efficiency:} Determine if the current search is efficient and terminate early if not. In contrast, the existing state-of-the-art technique selects a random path from the full list of differentials which has a higher degree of variance and is less efficient. }
    \label{fig:flow_contribution}
\end{figure*}

While the results presented by \cite{dwivedi2023security} show considerable improvements over previous techniques, the authors identify that the algorithm is inefficient for large-state block cyphers. While the technique improves performance over previous methods considerably, we have identified that the use of \textit{simple random sampling} hampers the search performance. Although simple random sampling of the entire population is easy to implement, a high level of variance within the full distribution of differentials can impede performance and reduce efficiency. Although Monte-Carlo search typically results in improved efficiencies, stratified sampling \cite{etikan2017sampling} has been identified as a method of variance reduction to improve estimate precision, improve efficiency and reduce time in Monte-Carlo search \cite{press1990recursive, sarndal2003model, taverniers2020estimation, napolitano2018lightning, shields2015refined}. While sampling can produce improved results, the random nature of Monte Carlo Search (MCS) can still lead to inefficient outcomes. In situations where the search has become inefficient, the authors of \cite{lanctot2014monte} noted that early terminations can improve the performance of the heuristic. Subsequently, the following thought-provoking research questions (RQ) arise:
\begin{itemize}
    \item \textit{RQ1: How do the changes to the variance within the sample differentials influence the efficiency of the NMCS algorithm?}
    \item \textit{RQ2: What impact does VISTA-CRYPT sampling have on the efficiency of differential cryptanalysis for SIMON $32$ and SIMECK $32$ cyphers in terms of duration and number of iterations?}
    \item \textit{RQ3: What are the effects of early termination of inefficient searches of the quota sample between experiments?}
\end{itemize}

Having identified the limitations of the current state-of-the-art solution, this work introduces a new technique for differential cryptanalysis called VISTA-CRYPT. Our technique employs a form of stratified sampling, known as quota sampling \cite{sharma2017pros}, of the differential paths, reducing variance and population size. Through the application of VISTA-CRYPT we are able to demonstrate improved efficiencies with results that can be measured in seconds on small block sizes, rather than minutes. We expand on our contribution by presenting a detailed analysis of the results of VISTA-CRYPT highlighting efficiency gains compared to the existing state-of-the-art results. A concise illustration of our contribution is presented in Figure \ref{fig:flow_contribution} comparing the existing state-of-the-art work with our own technique. To summarise, our key contributions are as follows:

\begin{itemize}
    \item Established reductions in differential variance between the existing state-of-the-art and VISTA-CRYPT, addressing RQ1.
    \item Demonstrated efficiency gains to the existing state-of-the-art through the introduction of quota sampling and reduced variance, addressing RQ2.
    \item Presented a detailed analysis of the VISTA-CRYPT technique by identifying two key performance metrics and contrasting the results to the existing state-of-the-art technique. The analysis underscores two key efficiency objectives: $1$) A reduction in the number of iterations required to reach the target hamming weight. $2$) A reduction in the total time required to execute the algorithm.
    \item Performed a preliminary graph-based analysis of the output differentials which allows for the identification of relationships within the data and future research opportunities to further enhance the performance of differential cryptanalysis.
\end{itemize}

The remaining sections of this paper are structured as follows: Section \ref{Preliminaries} describes the background information of the domain necessary for this article. Section \ref{Sec:Relatedwork} describes the state of the art in existing lightweight encryption algorithms and presents a brief literature review of related works. Section \ref{Sec:SystemModel} introduces the proposed system model to improve algorithm efficiency. Section \ref{Sec:Methods} elaborates on the methodology adopted for this article and describes the algorithms used in VISTA-CRYPT. The results and findings of this investigation are summarised in Section \ref{Sec:Results} and the findings and future work are discussed in Section \ref{Sec:Discussion}. A conclusion is presented in Section \ref{Sec:Conclusion} of the article.

\section{Preliminaries }\label{Preliminaries}

This section provides the reader with the foundational knowledge of the domain necessary for our state-of-the-art contribution. 
Readers are referred to Table \ref{Tab:Symbols} for a description of the symbols and notations used throughout this article.

\begin{table}
\caption{Symbols and notations}
\centering

%\resizebox{6cm}{!} {
\begin{tabular}{ll}
\hline
\textbf{Notation}      & \textbf{Description}  \\ \hline
$n$         & Number of bits \\ \hline
$m$         & The keyword size of either 2, 3, 4  \\ \hline
$\land$          & Bitwise AND \\ \hline
$\lor$          & Bitwise OR    \\ \hline
$\oplus$        & Bitwise exclusive OR (XOR) \\ \hline
$\boxplus$        & Modulo addition \\ \hline
$ \xrightarrow{n} $        & Right shift by $n$ bits \\ \hline
$ \xleftarrow{n}$        & Left shift by $n$ bits \\ \hline
$k_{(i)}$        & Key $i$ round \\ \hline
$\sigma^2$      & Population variance \\ \hline 
$N$             & Number of observations \\ \hline 
$x$             & Individual observations \\ \hline
$\mu$           & Population mean   \\ \hline
$S^2$           & Sample variance   \\ \hline 
$Y_i$           & Individual sample observation \\ \hline
$\overline{Y}$  & Sample mean \\ \hline
$p_{\beta}(b)$ & Probability density function of $\beta$   \\ \hline
\end{tabular}

\label{Tab:Symbols}
\end{table}

\subsection{SIMON Cypher }\label{DescriptionSimonCypher}

The SIMON cypher comprises five variants of $2n$-bit states, where $n$ denotes the word size. With $n = 16, 24, 32, 48, 64$, supporting a block size of $32, 48, 64, 96$ and $128$ bits. The key sizes of SIMON are composed of $m \times n$ bit words, where $m = 2, 3, 4$ and is based on the size of $n$. The size of $m$ must follow the following rules \cite{beaulieu2013simon}:
\begin{itemize}
    \item $m$ must be $4$ if $n$ equals $16$.
    \item $m$ may be $3, 4$ if $n$ equals $24$ or $32$.
    \item $m$ may be $2, 3$ if $n$ equals $48$.
    \item $m$ may be $2, 3, 4$ if $n$ equals $64$.
\end{itemize}
Using these parameters, SIMON can be represented as follows. SIMON $2 n / mn$ has a block size of $2n$ bits and a key size of $m \times n$ bits \cite{beaulieu2013simon}. As an example, SIMON$32$/$64$ refers to the version of SIMON acting on $32$-bit plaintext blocks and using a $64$-bit key comprising $4(m) \times 16(n)$, where $n$ is the word size.

\subsubsection{Round functions }\label{Subsec:roundFunctions}

As with all Feistel block cyphers, SIMON and SIMECK utilise round functions for the encryption and decryption process. According to the NSA \cite{beaulieu2013simon}, the SIMON round function makes use of bitwise XOR $(\oplus)$, bitwise AND ($\land$) and left circular shift $(\xleftarrow{i})$ by $i$ bits. The SIMON round function shown in Figure \ref{fig:simon_round_function} can be defined as:

\begin{equation}
    f(x) = \left( \xleftarrow{1} x \right) \land \left( \xleftarrow{8} x \right) \oplus \left( \xleftarrow{2} x \right) .
\end{equation}

\subsection{SIMECK Cypher }\label{DescriptionSimeckCypher}

SIMECK, on the other hand, only supports $3$ variants with $2n$-bit states with word size $n = 16, 24,  32$. The SIMECK block sizes are therefore $32$, $48$ and $64$ bits respectively. Table \ref{tab:simon_speck_simeck_parameters} illustrates the parameters of each of the encryption algorithms \cite{beaulieu2013simon}.

\begin{figure*}[t]
     \centering
     \begin{subfigure}[b]{.45\textwidth}
         \centering
         \includegraphics[width=1\textwidth]{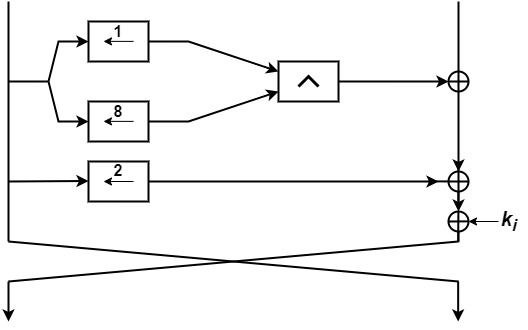}
         \caption{SIMON round function}
         \label{fig:simon_round_function}
     \end{subfigure}%
     \hfill
     \begin{subfigure}[b]{.45\textwidth}
         \centering
         \includegraphics[width=1\textwidth]{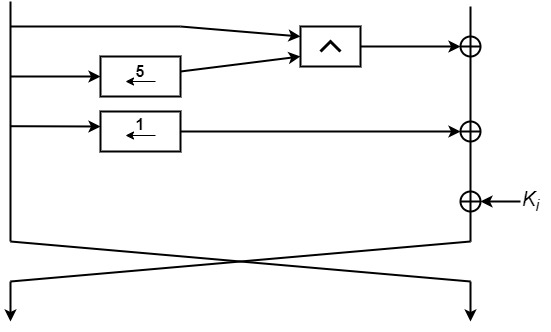}
         \caption{SIMECK round function}
         \label{fig:simeck_round_function}
     \end{subfigure}
        \caption{SIMON and SIMECK round functions}
        \label{fig:sinom_simeck_round_functions}
\end{figure*}

The SIMECK cypher has been designed to incorporate features of both the SIMON and SPECK cyphers, and as a result, the round function of SIMECK is similar to that of SIMON. The round function uses three operations, bitwise AND ($\land$), bitwise XOR ($\oplus$) and left circular shifts ($\xleftarrow{i}$) \cite{yang2015simeck}. The round function is illustrated in Figure \ref{fig:simeck_round_function} and is defined as:

\begin{equation}
    f(x) = \left(x \land (\xleftarrow{5} x )\right) \oplus (\xleftarrow{1} x ).
\end{equation}

\begin{table}
\centering
\small
\caption{SIMON, SPECK and SIMECK parameters}
%\begin{tabular}{llllll}
\begin{tabular}{|l|p{0.8cm}|p{0.8cm}|p{0.8cm}|p{0.8cm}|p{0.8cm}|p{0.8cm}|}
\hline
Variant  & Block \newline Size ($2n$) & Word Size ($n$) & Key Size $mn$ & Key Words & Rounds \\ \hline
SIMON32  & 32              & 16            & 64          & 4         & 32     \\ \hline
SIMON48  & 48              & 24            & 72          & 3         & 36     \\ \hline
         &                 &               & 96          & 4         & 36     \\ \hline
SIMON64  & 64              & 32            & 96          & 3         & 42     \\ \hline
         &                 &               & 128         & 4         & 44     \\ \hline
SIMON96  & 96              & 48            & 96          & 2         & 52     \\ \hline
         &                 &               & 144         & 3         & 54     \\ \hline
SIMON128 & 128             & 64            & 128         & 2         & 68     \\ \hline
         &                 &               & 192         & 3         & 69     \\ \hline
         &                 &               & 256         & 4         & 72     \\ \hline
% SPECK32  & 32              & 16            & 64          & 4         & 22     \\ \hline
% SPECK48  & 48              & 24            & 72          & 3         & 22     \\ \hline
%          &                 &               & 96          & 4         & 23     \\ \hline
% SPECK64  & 64              & 32            & 96          & 3         & 26     \\ \hline
%          &                 &               & 128         & 4         & 27     \\ \hline
% SPECK96  & 96              & 48            & 96          & 2         & 28     \\ \hline
%          &                 &               & 144         & 3         & 29     \\ \hline
% SPECK128 & 128             & 64            & 128         & 2         & 32     \\ \hline
%          &                 &               & 192         & 3         & 33     \\ \hline
%          &                 &               & 256         & 4         & 34     \\ \hline
SIMECK32 & 32              & 16            & 64          & 4         & 32     \\ \hline
SIMECK48 & 48              & 24            & 96          & 4         & 36     \\ \hline
SIMECK64 & 64              & 32            & 128         & 4         & 44     \\ \hline
\end{tabular}%

\label{tab:simon_speck_simeck_parameters}

\end{table}

\subsection{Differential cryptanalysis }\label{Subsec:DifferentialCryptanalysis}

Differential cryptanalysis is typically used to attack symmetric key algorithms and is an examination of how differences in an input affect differences in the output \cite{ biham2012differential}. In more simple terms, differential cryptanalysis searches for relationships between changes in the output from changes to the input. As \cite{easttom2021cryptanalysis} explains, by studying these changes in the output, it is possible to reveal some properties of the secret key. Differential cryptanalysis works by measuring the exclusive OR (XOR) difference between two values creating a characteristic that demonstrates that the specified differential, or change in the input leads to a particular differential, or change in the output. Developed in the late 1980s by \cite{biham1991differential} to decrypt the block cypher Fast data Encipherment Algorithm (FEAL), it has grown in popularity as a powerful tool for measuring changes throughout a cryptanalysis function.

\subsection{Calculating the XOR  differential probability of AND}

To determine the path of highest probability, we need to determine the XOR differential probability of AND. Both SIMON and SIMECK use bitwise AND components where the output is not proportional to the inputs. Throughout each round of the function, AND takes two inputs of which the output needs to be determined with high probability. The differential probability can be calculated with the following definitions \cite{abed2015differential}.

\textbf{Definition 1: }
The XOR differential probability of the logical AND operation measures the likelihood that when two XOR inputs of $(p, q)$ and  $(p', q')$ with corresponding intermediate XOR values (a, b), the XOR of the logical AND results in (c) matches a given value (c) across all possible input values ($p, q$) for $n$-bit binary numbers.

\begin{equation}
    xdp^\land(a, b \rightarrow c) = \frac{ \left| \{(p,q):f(p,q)\oplus f(p\oplus a, q \oplus b) = c \} \right| } {2^{-2n} },
    \label{eq:xor_differential_probability}
\end{equation}

\indent where $x$ is the differential probability following the AND operation and determines the probability that $x$ is either $0$ or $1$ and $n$ is the state size of the cypher. This illustrates that the probability of selecting the best differential path is near zero, and as such, will require more rounds to increase the probability.

\textbf{Definition 2: } Let the hamming weight function of $hw(\cdot$) and $a, b, c$ be of fixed $n-bit$ XOR differences such that:

% \begin{equation}
%     xdp^\land(a, b \rightarrow c) = \Biggl\{ {~~0~~~~~~~~~~~~~~~, \text{if  $a=b=0$} \atop 2^{-hw(a \lor b)}, otherwise}
% \end{equation}

\begin{equation}
xdp^\land(a, b \rightarrow c) = \left\{ 
\begin{array}{ll}
0, & \text{if } a=b=0 \\
2^{-hw(a \lor b)}, & \text{otherwise}
\end{array} 
\right.
\end{equation}

\begin{figure}[t]
    \centering
    \includegraphics[width=0.6\textwidth]{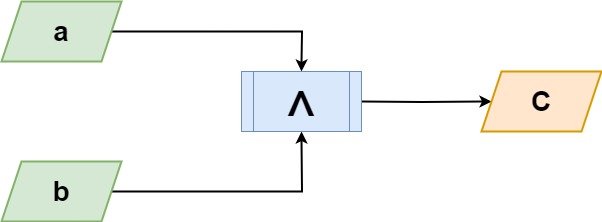}
    \caption{Transition through Bitwise AND}
    \label{fig:transition_bitwise_and}
\end{figure}

The authors of \cite{dwivedi2023security} describe the heuristic approach to find the differential path in the cypher as using a binary tree-like structure consisting of zeros and ones that is based on random sampling. To find the differential path, each point in the tree where there is a non-linear change in the output from the input is considered a decision point. The decision consists of two input differences, $a$ and $b$, where we need to find the output difference of $c$ with high probability, as illustrated in Figure \ref{fig:transition_bitwise_and}. This is then applied to the tree-like structure to determine the best path.

\begin{figure}[t]
    \centering
    \includegraphics[width=0.6\textwidth]{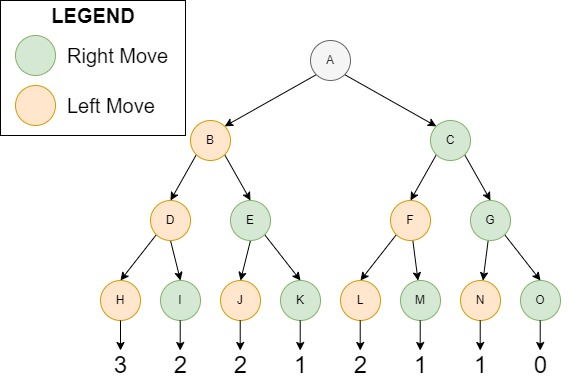}
    \caption{Heuristic nested tree search}
    \label{fig:NMCS_random_tree}
\end{figure}

Consider Figure \ref{fig:NMCS_random_tree} representing the tree-like structure mentioned above. Each move to the left from the parent node to an orange node has a cost of one and each move to the right to a green node has a cost of zero. The heuristic NMCS seeks to find the best path by reducing the total cost through random sampling at each node. Suppose the algorithm has a goal hamming weight of one or less, and at each node along the path there is a fifty per cent chance of taking either the left or right path. The first path explored is \textit{A, B, D, I} with a weight of $2$. The path and weight are saved as the current best path and weight. The next path explores \textit{A, B, E, J} with a weight of $2$. As this path has the same weight, the current best path is not updated. The third path explores \textit{A, B, E, K} with a weight of $1$. This path is then updated as the current best path, and as the goal weight has been reached the algorithm terminates. Although several paths exist in our example tree with a total weighted cost of one or less, the random path selection of the heuristic illustrates that it may not lead to the best path available. However, as demonstrated, suitable paths that lead to an acceptable result can be found through random investigation.

\subsection{Hamming weight }\label{Subsec:hammingWeight}

In differential cryptanalysis, hamming weight plays a crucial role in determining the extent of the changes in the differential process \cite{pestunov2014impact}. In Section \ref{Subsec:DifferentialCryptanalysis} above, we discuss the process of differential cryptanalysis and how it measures changes to the output from changes to the input. The changes are quantified using the hamming weight, which is the measure of non-zero bits in a binary string \cite{thompson1983error}. By calculating the hamming weight of the differential it is possible to characterise the differences between pairs of plaintexts and corresponding cyphertexts. A high hamming weight is an indicator of significant differences between the two texts, while a low hamming weight indicates fewer differences. As the differential cryptanalysis converges towards the desired result, the hamming weight reduces. The hamming weight is further used to determine the differential probability by calculating the probability of the input and output changes.

\subsection{Quota sampling}

\begin{figure}[t]
    \centering
    \includegraphics[width=0.6\columnwidth]{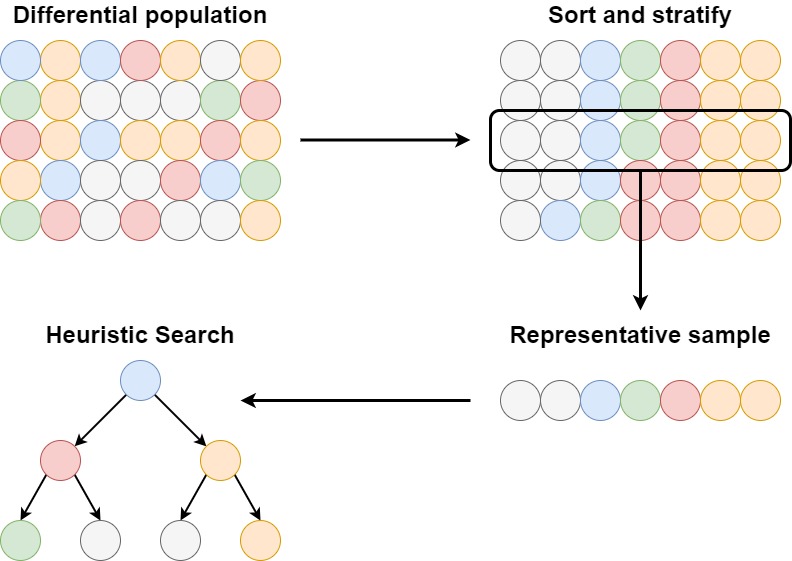}
    \caption{Heuristic NMCS with quota  sampling}
    \label{fig:strata_sample_flow}
\end{figure}

Earlier authors in \cite{sharma2017pros} describe quota sampling as the non-probability based equivalent of stratified sampling. Quota sampling involves dividing a population into smaller, homogenous groups based on a specific characteristic, and then selecting a proportionate sample from each group. An example of this process is illustrated in Figure \ref{fig:strata_sample_flow} which contains a population of different coloured cells. The population is first sorted into subgroups depending on their colour. They are then proportionally represented in the sample. As shown, the light grey and light orange subgroups contain two items each. The light blue, light green and pink subgroups each contain one. Our sample size is thus seven with two light grey, two light orange, one light blue, one light green and one pink which can be used to form the heuristic tree. Quota sampling serves two purposes. First, reducing the population size and secondly, reducing variance within the distribution. 

\section{Related work }\label{Sec:Relatedwork}

\begin{table}[t]
\centering
\tiny
\caption{Related Literature}
\label{tab:related-literature}
\resizebox{\textwidth}{!}{%
\begin{tabular}{|c|p{3cm}|p{4cm}|p{4cm}|}
\hline
Reference                       & Research Scope                                                                                  & Findings                                                                                                                               & Limitations                                                                                                                                                                 \\ \hline
\cite{dwivedi2023security}      & Analysis of differential   cryptanalysis on SIMON and SIMECK using nested Monet-Carlo search.   & The technique improved   efficiency and is an easy to implement solution for the differential   cryptanalysis of other block cyphers.  & The technique is still   inefficient on large state block cyphers and it uses random sampling which   produces different results each time.                                 \\ \hline
\cite{dwivedi2020security}      & The differential cryptanalysis   of ARX based cypher Chaskey using Monte-Carlo tree search.     & Heuristic search method produced   results that are significantly faster than an exhaustive search.                                    & The use of random sampling   produces different results each time.                                                                                                          \\ \hline
\cite{biryukov2015differential} & Analysis   of the differential properties of SIMON and SPECK.                                   & Applying new techniques of   automatic searching of SIMON and SPECK resulted in improved attack   efficiencies.                        & Predates the introduction of   SIMECK.                                                                                                                                      \\ \hline
\cite{dwivedi2018differential}  & Nested   Monte-Carlo Search algorithm to find a differential path in ARX cypher LEA.            & Discovered differential paths   for up to 12 rounds in reduced time and provided a reusable framework for   other avenues of research. & Limited scope of investigation   focusing research on only one family of cypher.                                                                                            \\ \hline
\cite{dwivedi2019differential}  & Differential cryptanalysis of   SPECK cypher.                                                   & Improved differential   cryptanalysis of SPECK using NMCS with a partial difference distribution   table.                              & The use of a partial difference   distribution table may result in missed values for a good differential path.   Additionally, the research is limited to the SPECK cypher. \\ \hline
\cite{dhar2018finding}          & Propose the adaption of Nested   Monte-Carlo Search to find differential trails in ARX cyphers. & Using NMCS provided similar   results to previous work by \cite{biryukov2015differential}, but with a   simpler implementation.        & Investigation is limited to   SPECK32 cypher.                                                                                                                               \\ \hline
\cite{abed2015differential}     & Differential cryptanalysis of   round-reduced SIMON and SPECK cyphers.                          & Early demonstration that up to   half the rounds of SIMON and SPECK can be attacked using differential   cryptanalysis.                & Pre-dates the introduction of   the SIMECK cypher.                                                                                                                          \\ \hline
\cite{napolitano2018lightning}  & Monte Carlo Search using   stratified sampling to assess lightning performance of power lines.  & Demonstrates a significant time   reduction over standard Monet Carlo methods.                                                         & Study was undertaken in an unrelated   field on standard Monte Carlo Search.                                                                                                    \\ \hline
\cite{taverniers2020estimation} & Study on distributions via   multilevel Monte Carlo using stratified sampling.                  & Reduced computational costs and   increased efficiencies.                                                                              & Additional research is required for   when the probability of failure of the search is required.                                                                                \\ \hline
\end{tabular}%
}
\end{table}

This section presents a brief discussion and literature review of related work on differential cryptanalysis which has been summarised in Table \ref{tab:related-literature}.

The review will begin with a discussion on the cryptanalysis of the lightweight cyphers and previous work on similar block cyphers. The discussion will move to the existing state-of-the-art methods, their results and identified areas of additional research. We will conclude this section with a discussion of the use of stratified sampling in MCS and the potential benefits of the use of quota sampling, which is the non-probability based equivalent of stratified sampling \cite{sharma2017pros}, in differential cryptanalysis. To the best of our knowledge, no prior work exists on the use of quota sampling to improve the efficiency of NMCS differential cryptanalysis.

With the introduction of lightweight cyphers such as SIMON and SPECK in 2013, researchers rapidly began investigating efficient methods of cryptanalysis of the latest cyphers. Within two years of the introduction of SIMON and SPECK by the NSA, the authors of \cite{biryukov2015differential} conducted an analysis of the differential properties of both SIMON and SPECK, which was an extension of their previous work on automatic searches of differential trails on addition, rotate, XOR (ARX) cyphers \cite{biryukov2014automatic}. In their study, \cite{biryukov2015differential} documented that recent discoveries in automated search techniques resulted in better differential trails for SIMON $32$, SIMON $48$ and SIMON $64$, as well as SPECK $32$, SPECK $48$ and SPECK $64$ with one additional round exploited than with previous efforts. To compliment the work of \cite{biryukov2015differential}, the authors of \cite{abed2015differential} demonstrated that half of the rounds in both SIMON and SPECK can be successfully attacked using differential cryptanalysis. Although the early research by \cite{abed2015differential} and \cite{biryukov2015differential} resulted in successful attacks on a reduced number of rounds of SIMON and SPECK using differential cryptanalysis, the research demonstrated the effectiveness of differential cryptanalysis in attacking novel lightweight block cyphers.

The use of MCS as a tool for differential cryptanalysis of ARX cyphers was proposed by the authors of \cite{dhar2018finding} in 2018 when they demonstrated that the novel method was capable of replicating results similar to those of \cite{biryukov2015differential} against the SPECK $32$ cypher. MCS was later used to demonstrate its effectiveness against ARX cypher Light Encryption Algorithm (LEA) when the authors of \cite{dwivedi2018differential} were able to increase to fourteen the number of rounds successfully attacked using their proposed method. However, while existing research has focused on SPECK, SIMON and other ARX cyphers such as LEA, investigations into the effectiveness of MCS against SIMECK remained untested.

Building on earlier research by the authors of \cite{dwivedi2020security, dwivedi2018differential, dwivedi2019differential} and \cite{dhar2018finding}, the authors of \cite{dwivedi2023security} proposed a state-of-the-art heuristic search method as an efficient way to conduct a differential cryptanalysis attack on SIMON and SIMECK cyphers without the need for high computational servers or clusters. The authors of \cite{dwivedi2023security} demonstrated that commercially available consumer devices, such as a laptop, can be used to conduct an effective round-reduced attack on the SIMON and SIMECK cyphers in a matter of minutes, instead of hours or days. However, while their proposed method offers improvements in efficiency over existing methods, certain aspects of the approach degrade the performance. Particularly, the use of simple random searching to determine the path is prone to a high degree of variance within the differential distribution, resulting in an inefficient search.  Further, the authors note that their method remains inefficient for large block-size cyphers. While the authors proposed a method for improving efficiency, by splitting the algorithm in half and conducting an analysis from the middle, one moving forward and the other backwards, this solution does not address the problem of simple random sampling from a large population.

The use of stratified sampling to improve MCS efficiency dates back over $30$ years. One of the first proposals of using stratified sampling to improve MCS was introduced by \cite{press1990recursive} in 1990 which, similar to \cite{shields2015refined}, was a modified version of stratified sampling. Although predominantly concerned with physics, it presented one of the first instances of stratified sampling to improve the results of MCS. A year later, the authors of \cite{marnay1991effectiveness} presented the results of their investigation of stratified sampling on MCS. In their study, the authors noted that the use of random sampling in MCS introduces a degree of imprecision, also known as sampling variance, into the output of the algorithm. Although they illustrate that variance reduction techniques can reduce variance and improve precision, they note that the improvements can vary due to rare events. More recently, \cite{napolitano2018lightning} illustrated a computational time reduction of over $75\%$ by introducing stratified sampling in their study. Further, a study by \cite{taverniers2020estimation} supported the findings of \cite{shields2015refined} and \cite{napolitano2018lightning} that the use of stratified sampling in MCS results in variance reduction and reduced computational costs. Although the research of \cite{taverniers2020estimation}, \cite{napolitano2018lightning} and \cite{shields2015refined} was unrelated to differential cryptanalysis, the evidence of the significance of stratified sampling, particularly when applied to a large search space to improve MCS efficiency, is robust. Although existing work has demonstrated the effectiveness of employing stratified sampling in MCS, it has yet to be implemented in cryptanalysis.

\section{Problem Formulation}\label{Sec:SystemModel}

\begin{figure*}[t]
    \centering
    \includegraphics[width=0.9\textwidth]{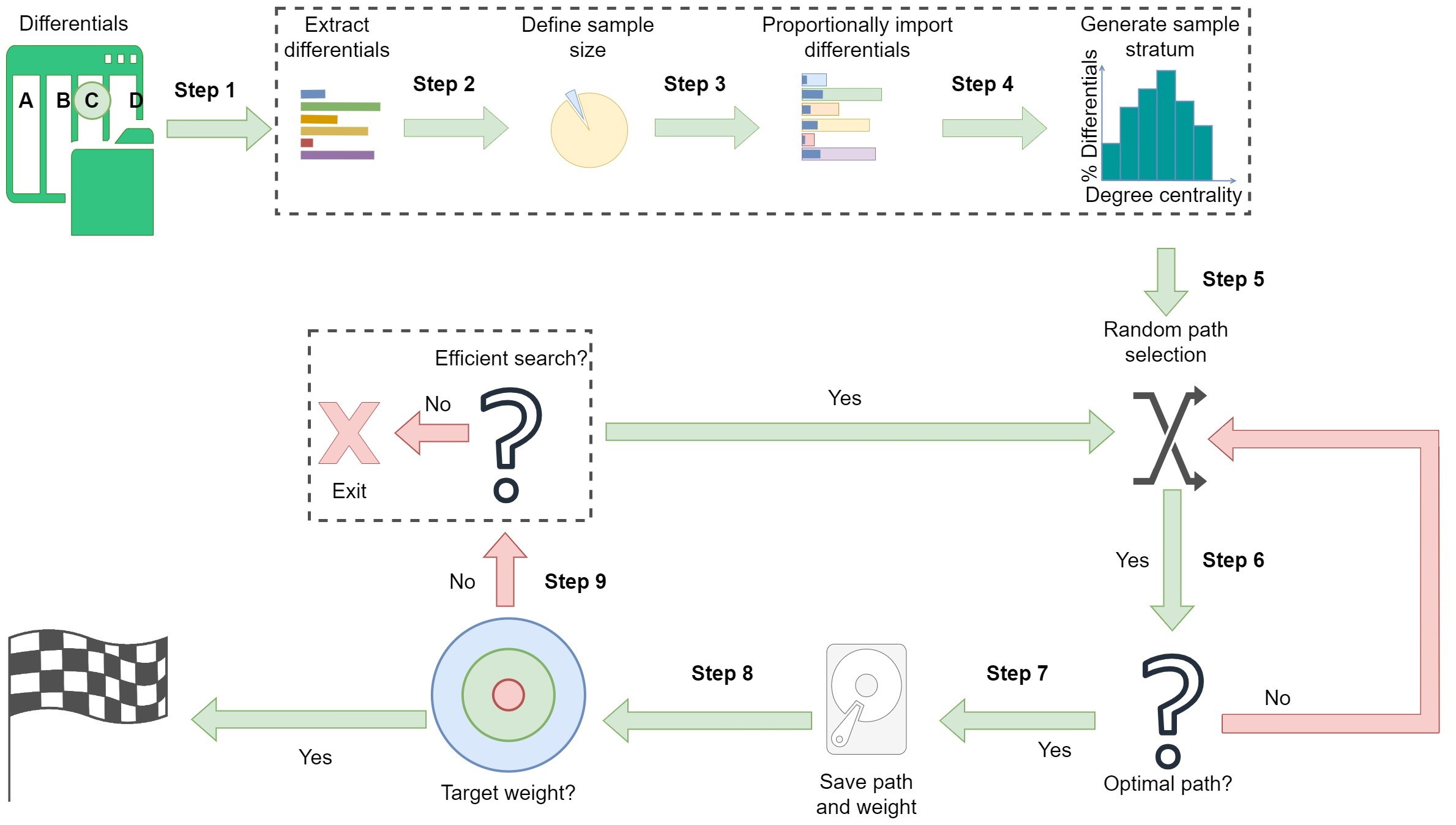}
    \caption{Our technique with our contributions outlined with a dashed line: $1$) \textit{Extract output differentials:} Extract output differentials (C) from lists that also contain left input (A), right input (B) and weight (D). $2$) \textit{Define proportional sample size:} Define the proportion of differentials to use in the
sample. $3$) \textit{Differential extraction:} Proportionally extract the differentials with a minimum of one of each type to the sample stratum. $4$) \textit{Sample generation:} A sample based on proportional stratification is generated. $5$) \textit{Random path selection:} A random path from the sample is chosen. $6$) \textit{Decision point - Is current path optimal: } Perform a check to determine if the current path is optimal. If yes then proceed, if not then return to step ($5$). $7$) \textit{Save path and weight:} If the current path is optimal save the current path and weight. $8$) \textit{Test - Target hamming weight reached:} Determine if target hamming weight has been attained. If the target weight has been reached then complete the search, if not then proceed. $9$) \textit{Decision point - Is search efficient:} Perform a check to determine if the current search is efficient. If yes then return to step ($5$), if not then terminate the search. }
    \label{fig:VISTA-CRYPT_flow_diagram}
\end{figure*}

As previously discussed, the discovery of an optimal path through random discovery is small, and as such, will require significantly more iterations to discover an optimal path. More iterations will result in an increase in computational resources and the time required to undertake an attack. The required code iterations and time variables will continue to grow exponentially as the search space increases with the key size. To counter the complexity of finding the most appropriate path in the shortest possible time, we propose the use of quota sampling with NMCS, which we call VISTA-CRYPT, coupled with the identification and early termination of inefficient searches. We have visualised the algorithm process in Figure \ref{fig:VISTA-CRYPT_flow_diagram}, with our contributions highlighted by the dashed borders.

As identified above in Section \ref{Sec:Introduction}, the existing NMCS algorithm presented by \cite{dwivedi2023security} offers an efficient means of iterating through differential paths to determine the fastest possible result of a differential attack on the SIMON and SIMECK cyphers. However, the recursive algorithm responsible for determining the shortest path makes use of simple random population sampling. With a large number of differentials with high variance to randomly select, the probability of randomly selecting a path that returns a better result than during the previous iteration is low. Although vastly more efficient than previous attempts at attacking SIMON and SIMECK using differential cryptanalysis, simple random sampling from the entire differential distribution is at a disadvantage when applied to path selection in an NMCS algorithm, particularly as the search space grows. As illustrated in Figure \ref{fig:strata_sample_flow}, by applying quota sampling to the full differential distribution there is a significantly smaller search space, reducing variance and increasing the probability of randomly discovering an optimal path. Although quota sampling improves efficiency, the heuristic technique of randomly searching differentials can produce inefficient searches that require additional iterations and longer processing time. To address experiments where the search becomes inefficient, the upper quartile of the number of iterations of the existing state-of-the-art has been identified as a termination point to conserve resources and maintain efficiency. These inefficient searches are then filtered out of the data for analysis.

\section{Methodology }\label{Sec:Methods}

In this section, we present our technique to generate quota samples from a list of output differentials for the SIMON and SIMECK cyphers. Figure \ref{fig:flow_contribution} summarises our sampling procedure and early termination with a comparison to the existing state-of-the-art method, with the full system model illustrated in Figure \ref{fig:VISTA-CRYPT_flow_diagram}.

In the work by the authors of \cite{dwivedi2023security}, the timings of their script were the primary factor measuring the overall performance of their cryptanalysis technique. In undertaking the measurements of time for their script, the authors used a standard Mac operating system laptop with $2.3$ GHz dual-core processor and $8$ GB RAM. While the authors do not state how the time was recorded, the provided code in their GitHub repository offers no time-recording functions or libraries. It could therefore be reasonably assumed that the timings were conducted manually or by the use of IPython magic commands $\%\%time$ and $\%\%timeit$ \cite{ipythonBuiltinMagic}. However, the magic commands are still absent from the GitHub files raising concerns as to how the times were observed. This approach raises several questions about reproducibility as a different device with different specifications will yield different results, particularly with timings. To counter these limitations, this paper introduced two variables for accurate comparative analysis.

The first measurement added is time recording introducing a reliable variable devoid of human error. Although the addition of time recording guarantees the accuracy of the results, it is still heavily dependent on the device running the script and can vary depending on the resources available. Nevertheless, as the original work used this metric for evaluation of their method, it can be used as a measure to determine the effectiveness of our modifications to the original script and concept. Furthermore, this addition will allow for comparisons across devices of varying processing power without bias. To complement the addition of accurate time recording, our work analyses the total number of recursive function iterations during the differential cryptanalysis. As noted by \cite{sidiroglou2011managing}, loops, particularly in Monte Carlo simulations, can result in inefficient operations. By recording the total number of iterations and reducing the total number through application enhancements, we can improve overall simulation efficiency while additionally providing a practical tool for performance analysis. However, due to the nature of random path selection, results will vary on each execution of the code. To gain an approximation of overall performance, we conducted $193$ experiments on each implementation with the results of the experiments shown in Tables \ref{tab:summary-comparison}, \ref{tab:summary-iterations} and \ref{tab:simeck32_analysis_28_weight}. Additional tables and plots from our analysis are discussed in Section \ref{Sec:Results}.

\subsection{Description of algorithm and sampling methods}

As highlighted above, the concept of quota sampling ensures that the sample contains a proportional representation of the entire population. While tools exist to automatically stratify the population depending on a defined number of strata and characteristics, the automated process can result in differentials with fewer presentations in the dataset being absent in some samples, or over represented in others. With simple stratified sampling techniques, problems can arise that can impede performance. Suppose a stratum is defined consisting of $40$ strata, yet one differential within the entire population is only present $12$ times. It is not possible to distribute that differential equally between all $40$ samples. While many experiments may conclude successfully in an efficient timeframe, should a differential be required and not present in the sample, it can result in longer processing times with many more iterations to converge on the ideal hamming weight. Rather than defining samples based on automatic stratification and then selecting a random stratum to experiment with differential cryptanalysis, our process employs quota sampling and defines the constraints of the stratum more rigidly to ensure a representative sample while reducing variance. Following are our quota rules for defining a sample:

\begin{itemize}
    \item \textit{Define the sample size:} For our analysis, we have defined a sample that is based on $5 \%$ of each type of nominal value, which represents $5 \%$ of the entire population size. This ensures that the sample is small and representative, yet contains enough data to provide reliable results.
    \item \textit{Ensure the sample contains at least one of each differential:} As some nominal differentials within the population have a small representation, a $5 \%$ rule would exclude many differentials from the sample. By guaranteeing that the sample contains at least one of each differential, we can ensure that the sample is representative of the entire population.
\end{itemize}

While our sampling method is not a true derivative of stratified sampling, it contains features representative of the technique and is considered a non-probability based equivalent to stratified sampling \cite{mujere2016sampling, sharma2017pros}. While it additionally makes use of an aspect of simple random sampling from a complete enumeration survey, such as $5\%$ of the population, the uniformed representation of $5\%$ of all nominal types of observation from the full distribution maintains the normality of the distribution with respect to the full population. While true stratification may group by ordinal, nominal or other data types, resulting in some stratum missing key observations, our approach seeks to make use of a representative sample, and in doing so reduces the size of the dataset and variance within the data. This process is imperative when dealing with a large list of differentials and a heuristic method of exploration. A poorly defined sample can lead to undesirable results while the heuristic searches for a differential that may not be included in the sample. This can lead to excessive processing times and unnecessary code iterations. The number of samples to take from the population of $n$ is given as:

\begin{equation} 
    n = \text{max} \left( 1, \left \lceil {\left(\frac{x}{100} \times N_h\right)} \right \rceil \right),
    \label{sample_eq}
\end{equation}
\indent where $x$ is the sample size of the stratum, and $N_h$ is the total number of observations of the $h^{th}$ type in the population, with at least one of each $h^{th}$ discrete value.

\subsection{Early termination}

As discussed earlier in this section and identified by \cite{sidiroglou2011managing}, successive loops in an NMCS can prove to be inefficient, and effectively managing loops that can constrain device functionality is a priority. When dealing with heuristic search in large state spaces, the random nature of the discovery can result in prolonged chains of processes that consume both resources and time. Due to the constrained resources of IoT devices, it is necessary to limit the time and resources that an algorithm consumes while undertaking differential cryptanalysis. An algorithm that continues to loop through differentials searching for the optimal path could cause the device to malfunction. In addressing this concern, we have introduced an early termination rule to the algorithm where the algorithm will terminate the search process once it reaches a predefined number of iterations. Through an analysis of the existing state-of-the-art results, we have based the maximum number of iterations as the upper quartile of the number of iterations for the existing state-of-the-art. When the algorithm has reached that number of iterations during the search, the algorithm will terminate and record the number of iterations, time, and current hamming weight for that experiment.

\subsection{Calculating results with high probability} 

Although our sampling method reduces the sample size and variance within the data, the randomness of the heuristic can produce different results with each experiment. Although not unexpected with a MCS, too few experiments can produce inconsistent results. To identify the number of simulations required for the creation of a synthetic dataset for comparative analysis, we must first conduct preliminary simulations to gain an estimation of the standard deviation ($\sigma$) within the sample. To gain a reliable standard deviation, we conducted $50$ simulations. The standard deviation from the preliminary simulation is then used to calculate the Standard Error (SE) that is defined as:

\begin{equation}
    SE =  \frac{\sigma}{\sqrt{n}},
\end{equation}
\indent where $\sigma$ represents the standard deviation defined from the preliminary simulations and $n$ is the number of simulations performed. We now define the Margin of Error (ME) for the $95\%$ confidence interval as:
\begin{equation}
    ME =  Z \times SE,
\end{equation}
\indent where $Z$ represents the $95\%$ confidence level ($1.96$).

We can solve for $n$, the number of NMCS required for our analysis with $95\%$ confidence as:
\begin{equation}
    n =  \left(\frac{Z \times \sigma}{ME} \right)^2,
\end{equation}
\indent where $Z$ is the confidence level, $\sigma$ represents the standard deviation and $ME$ represents the margin of error. This gives us:

\begin{align*}
    n &=  \left(\frac{1.95 \times 1.80558}{0.50048} \right)^2, \\
    &= 193.
\end{align*}

We calculate that a total of $193$ simulations will need to be performed which will produce results with high probability and account for the random nature of the heuristic.

\subsection{Experimental code}
Our implementation of differential cryptanalysis has been executed on SIMON $32$ and SIMECK $32$ block cyphers. To allow for a comprehensive comparison of the existing implementation of NMCS and our proposed enhancements, the following code adjustments have been applied to all sets of Python code, including those provided by \cite{dwivedi2023security} for comparative analysis: $1$) Inserted missing variables. $2$) Defining new variables for target hamming weight and iterations. $3$) Addition of timing variables to record algorithm timing. $4$) Iterable \textit{for loop} that executes the code $193$ times. $5$) Variables and functions to save the results to a comma-separated values (CSV) file. Beyond these corrections, we have applied code comments to enhance the readability of the code, which has no bearing on performance. 

\begin{algorithm*}
	\caption{Define Sample} 
    \label{Alg:DefineSample}
    \hspace*{\algorithmicindent} \textbf{Input}: Output differentials \\
    \hspace*{\algorithmicindent} \textbf{Output}: Sample of differentials
	\begin{algorithmic}[1]
    \Function{defineSample}{}
        \Comment{Using equation \ref{sample_eq}}
        \State $output\_differentials =$ List of  $output\_differentials$
        \Comment{Import the full list of output differentials} \newline
            \hspace*{1em}  for each $item$ in $highway\_list output differentials$ 
            \Comment{Loop through list of output differentials}
            
        \State \hspace*{1em} $item\_counts =$ Each item in $output\_differentials$
        \Comment{Count occurrence of each item in output differential}
        
        \State $final\_list = []$
        \Comment{Define the final list for the sample of differentials}
        \For{$item, count$ in $item\_counts$}
            \Comment{Loop through the count of items}
            \State $count\_to\_add = \lceil count \cdot 0.05 \rceil$ 
            \Comment{Include a 5\% sample with at least 1 of each}
            \State Append $item$to $final\_list$ $count\_to\_add$ times
            \Comment{Append each item to the final sample list}
        \EndFor
       
        \Return $final\_list$
        \Comment{Return the final sample list}
    \EndFunction    
	\end{algorithmic} 
\end{algorithm*}

\subsubsection{Construct the sample} 

Algorithm \ref{Alg:DefineSample} constructs a quota sample based on the principle of stratification. While typical stratification methods construct multiple samples and select one of the samples from the strata for analysis, in our approach we construct one sample ensuring proportional representation of all differentials with at least one of each differential present in the sample. By ensuring that all differentials are present in the sample, we avoid challenges that may arise if a differential is required, but not present, resulting in additional iterations and prolonged run time.

\begin{algorithm*}
	\caption{Create a Random Differential Path from Sample} 
    \label{Alg:RandomPath}
    \hspace*{\algorithmicindent} \textbf{Input}: Sample differentials \\
    \hspace*{\algorithmicindent} \textbf{Output}: Path weight, current path
	\begin{algorithmic}[1]
    
    \Function{RandomPath}{$current\_round\_position$}
    \Comment{Random path function}
        \While{$current\_round\_position \neq last\_round$}
        \Comment{Loop while current round position $\neq$ last round position }
            \State calculate hamming weight of non-linear layer with \newline
            \hspace*{2.6em} two inputs and random output \newline
            \hspace*{2.6em} from \Call{SelectRandomStratum}{\textit{differentials}}
            \Comment{Select random differential from sample defined in Algorithm \ref{Alg:DefineSample}}
        \EndWhile   \newline 
        \hspace*{1.2em} \Return $weight, path$
        \Comment{Return weight and current path}
    \EndFunction    
	\end{algorithmic} 
\end{algorithm*}

\subsubsection{Random differential path}

While the original version of Algorithm \ref{Alg:RandomPath} took a simple random sample of the entire output differential, our modification takes a simple random sample of the output differential from the sample defined in Algorithm \ref{Alg:DefineSample}. 

\begin{algorithm*}
	\caption{Nested Recursive Function} 
    \label{Alg:NestedRecursiveFunction}
    \hspace*{\algorithmicindent} \textbf{Input}: Weight, path \\
    \hspace*{\algorithmicindent} \textbf{Output}: Best weight, current round position
	\begin{algorithmic}[1]
    \Function{Nested}{$current\_round\_position$}
    \Comment{Recursive function}
        \While{$current\_round\_position \neq last\_round$} 
            \State \textit{weight, path$ = $RandomPath(current\_round\_position)}
            \Comment{Call weight and path from Algorithm \ref{Alg:RandomPath}}
            \If{$weight < best\_weight$  }
            \Comment{Check if current weight < best weight}
                \State $best\_path = path$
                \Comment{Update best path with current path}
                \State $best\_weight = weight$
                \Comment{Update best weight with current weight}
            \EndIf
            \State update current\_round\_position \newline
            \hspace*{2.6em} and go one level down by following best\_path
            \Comment{Update current round position}

            \If{$current\_round\_position \neq last\_round$  }
            \Comment{Check if the current round position $\neq$ the last round position} \newline
            \hspace*{4.2em}    \Call{Nested}{$current\_round\_position$}
            \Comment{Call recursive function}
            \EndIf         
        \EndWhile
    \EndFunction    
	\end{algorithmic} 
\end{algorithm*}

\subsubsection{Nested recursive function}

As described by \cite{dwivedi2023security}, Algorithm \ref{Alg:NestedRecursiveFunction} is a recursive function that calls itself at each level of the cypher round until it reaches the last round. This function retains two global variables called \textit{best\_path} and \textit{best\_weight}, which store the values of the lowest hamming weight in the \textit{best\_weight} variable and the current best path in the \textit{best\_path} variable. The \textit{best\_path} variable is initialised as an empty list and the \textit{best\_weight} is initialised with a large value with the goal of reducing the hamming weight.

\begin{algorithm*}
	\caption{Iterative Calls to Nested Function} 
    \label{Alg:IterativeCalls}
	\begin{algorithmic}[1]
    \State $best\_weight = 999$ 
    \Comment{Define best weight variable to a high initial value}
    \State $current\_round\_position =1$
    \Comment{Define current round position variable to 1}
    \State $last\_round = cypher\_round$
    \Comment{Define last round variable to cypher round}
    \While{ $i < max\_iterations$ and $best\_weight > target\_weight$} 
    \Comment{Iterative conditional calls}
    \newline
        \hspace*{1.2em} \Call{NESTED}{$current\_round\_position$}
        \Comment{Call Algorithm \ref{Alg:NestedRecursiveFunction}}
    \EndWhile
	\end{algorithmic} 
\end{algorithm*}

\subsubsection{Iterative calls to nested function}

Algorithm \ref{Alg:IterativeCalls} represents the main body of the code and is responsible for the iterative calls of Algorithms \ref{Alg:RandomPath} and \ref{Alg:NestedRecursiveFunction}. Algorithm \ref{Alg:IterativeCalls} can be called any number of times until the \textit{best\_weight} condition has been met, at which point the algorithm will terminate and return the \textit{best\_path} and \textit{best\_weight} values. If the number of iterations exceed the defined maximum number of iterations, the algorithm will terminate early and return the current \textit{best\_path} and \textit{best\_weight} values, even if they do not meet the target weight. This action is invoked as a method to maintain algorithm efficiency in situations where poor performance from the heuristic method impedes the outcome.

\section{Results and Analysis}\label{Sec:Results}

This section describes the results of the experimental implementation of the above algorithm enhancements described in Section \ref{Sec:Methods} on the NMCS differential cryptanalysis of the SIMON and SIMECK cyphers. As variance reduction has been identified as a contributing factor to improved performance of MCS, we will begin with an analysis of the variance reduction of quota sampling. Next, we present the results of the SIMON $32$ cryptanalysis using the one-way algorithm and then the SIMECK $32$ one-way cryptanalysis. This will be followed by the results of the two-way, forward and backwards, algorithm enhancements for both cyphers.

The experiments described in Section \ref{Sec:Methods} are deployed on a standard retail personal computer laptop with a Windows $10$ operating system, $12$th Gen Intel(R) Core(TM) i$7$-$1255$U $1.70$ GHz quad-core processor with $16.0$ GB RAM. All experiments are executed in IPython using Jupyter Notebook in the Microsoft Edge browser. As the algorithm employs random sampling, results between experiments will vary, so to address fluctuations in results, we have conducted $193$ experiments of both the existing state-of-the-art work presented by \cite{dwivedi2023security} and our proposed enhancements. The results of the experiments have been processed and analysed using IPython in Jupyter Notebook.

\subsection{Variance reduction analysis}

To quantify the effects of variance reduction and determine its extent of influence on the output differentials, we conducted a study of the variance within the full dataset of output differentials and the variance of the quota sample of output differentials. To determine the extent of variance reduction, we need to first establish the range of variance within the full output differentials. Using the Python NumPy library, the use of the \textit{var()} function allows for the computation of variance within the list of differentials. This function is executed on both the full distribution of differentials as well as the sampled distribution. From this, we can extrapolate the variance reduction as well as visualise the comparison of variance between the full and sampled lists. The population variance is represented as:

\begin{equation}
    \sigma^2 = \frac{1}{N} \sum_{i=1}^N  \left(x_i - \mu\right)^2 ,
    \label{eq:PVR}
\end{equation}
\indent where $N$ is the number of observations, $x$ is the individual observation and $\mu$ is the sample mean. The sample variance, not too dissimilar to the population variance, is represented as:
\begin{equation}
    S^2 = \frac{1}{n - 1} \sum_{i=1}^n \left(Y_i - \overline{Y} \right)^2,
    \label{eq:SVR}
\end{equation}
\indent where $N - 1$ represents the number of observations in the sample with $1$ degree of freedom to give an unbiased estimate of the population sample, $Y_i$ represents the individual sample observation and $\overline{Y}$ is the sample mean.

The reduction in variance can thus be calculated by:
\begin{equation}
    VR = \sigma^2 - S^2,
    \label{eq:VR}
\end{equation}
\indent where $\sigma^2$ represents the population variance and $S^2$ is the sample variance.

\begin{figure}[t]
    \centering
    \includegraphics[width=0.6\textwidth]{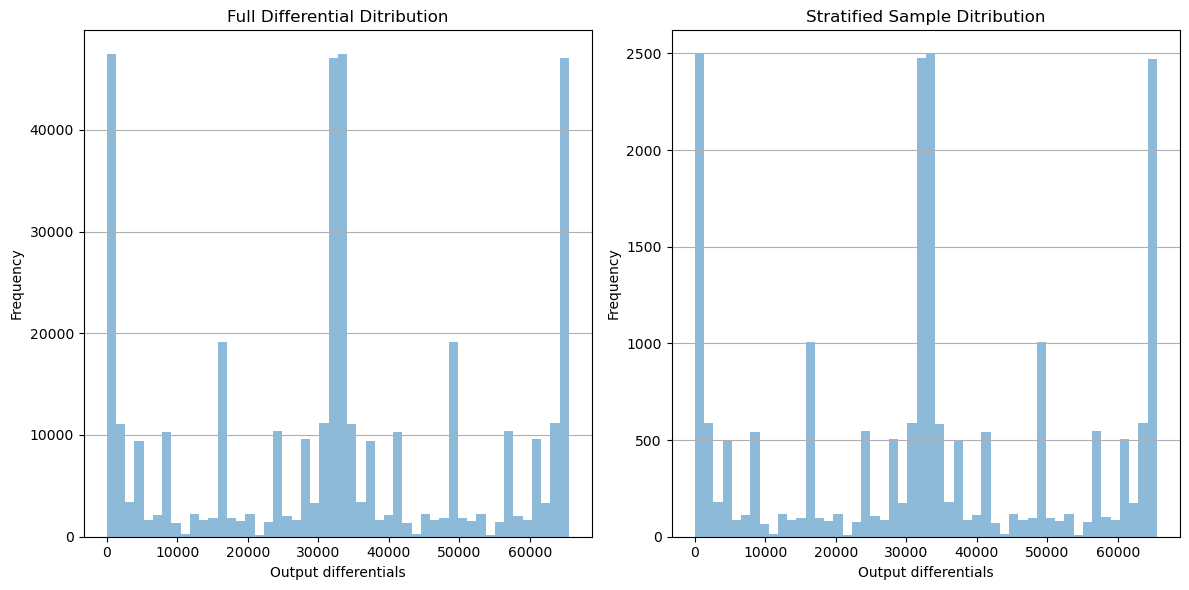}
    \caption{Differential distribution comparison of full differential distribution and sample distribution}
    \label{fig:differential_distribution_hist}
\end{figure}

\begin{figure}[t]
    \centering
    \includegraphics[width=0.6\textwidth]{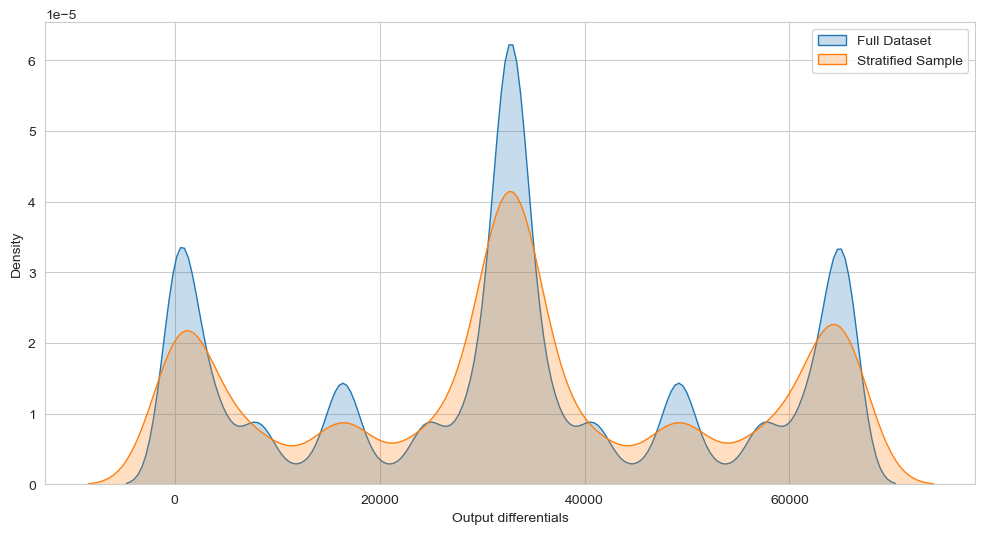}
    \caption{Differential distribution density comparison of full differential distribution and sample distribution}
    \label{fig:differential_distribution_density}
\end{figure}

On analysis, the variance of the full differential distribution is $440593216.09$ and the sample variance distribution is $439161728.57$, representing a reduction of $1431487.52$. As illustrated in Figure \ref{fig:differential_distribution_hist}, the sample variance of the data appears proportionally representative of the full differential distribution, however, with a significant reduction in the total number of differentials. This reduction and proportional representation is further highlighted in Figure \ref{fig:differential_distribution_density}, which allows for similar comparisons between the full and sample populations. However, as Figure \ref{fig:differential_distribution_density} clearly illustrates, the distribution of differentials is not normally distributed, as should be expected from a cypher, and has a multimodal distribution. As such, variance distribution across both the full population and sample will still exhibit a large spread from the mean, as highlighted above. Nevertheless, despite the persistence of a large variance from the mean, the sample demonstrated a significant reduction in variance which contributed to the efficiency gains in VISTA-CRYPT. With a clear reduction in sample variance evident, we can now proceed to the analysis of the application of VISTA-CRYPT in differential cryptanalysis.

\subsection{Data cleaning} 

To address the variations in the results due to the random nature of NMCS, it is necessary to clean the data, removing outliers and inconsistencies that can skew overall results. The process of data cleaning is conducted on all datasets to ensure conformity and data integrity. This process will begin by first removing all experiments with a weight not equal to the target weight defined in the code. For SIMON $32$ the target weight is $30$ and for SIMECK $32$ the target weight is $28$. It is important to note that with the introduction of early termination, the algorithm may at times terminate with a weight higher than the target weight, however, without early termination some experiments produce results below the target weight. Following the removal of weights below the target in the algorithm, we identify and remove outliers which have a tendency to skew data. To identify outliers within the datasets, we will use the interquartile range (IQR) method, which identifies data between the first quartile and third quartile. The data in this range focuses on the middle $50$ per cent around the median value and is thus not influenced by extremes in the data. The IQR rule is applied to both the duration and iteration variables. This process results in a collection of datasets devoid of outliers providing a more accurate interpretation of the data.

\subsection{Analysing results} 

Conducting an analysis of the sanitized data reveals promising results, as illustrated in Table \ref{tab:summary-comparison}. As shown, there is a considerable improvement in all metrics, with a marked reduction in variability between experiments as highlighted by the standard deviation. A comparison of time performance shows a mean reduction of $ 62.72\%$ and a median reduction of $60.59\%$, illustrating a significant improvement in the performance of VISTA-CRYPT over the existing state-of-the-art technique by \cite{dwivedi2023security}. The performance gains are further echoed by an analysis of the number of iterations by the algorithm, which can be viewed in Table \ref{tab:summary-iterations}. This, however, should be unsurprising, as the number of iterations of the code correlates with a causative effect on the duration of the algorithm execution. Although the reduction in iterations is not as pronounced as those of time, with the mean showing a decrease of $46.75$ per cent and a median decrease of $44.64$ per cent, it is evident that quota sampling improves algorithm efficiency.

\begin{table}
\centering
\tiny
\caption{SIMON32 differential cryptanalysis: Comparison of the existing state-of-the-art \cite{dwivedi2023security} and VISTA-CRYPT Algorithm Time (seconds)}
\label{tab:summary-comparison}
\resizebox{\columnwidth}{!}{%
\begin{tabular}{|l|l|l|}
\hline
       & Duration (s) existing state-of-the-art \cite{dwivedi2023security}  & VISTA-CRYPT Duration (s) \\ \hline
%Count  & 46       & 40          \\ \hline
Mean   & 34.28 & 12.78    \\ \hline
Standard Deviation    & 11.20 & 21.07    \\ \hline
Minimum    & 0.15 & 0.02    \\ \hline
25\%    & 4.76 & 1.95     \\ \hline
Median & 28.62 & 11.28    \\ \hline
75\%    & 53.59 & 19.43    \\ \hline
Maximum    & 122.24 & 39.00    \\ \hline
\end{tabular}%
}
\end{table}

\begin{table}
\tiny
\centering
\caption{SIMON32 differential cryptanalysis: Comparison of Number of Iterations of the existing state-of-the-art \cite{dwivedi2023security} and VISTA-CRYPT}
\label{tab:summary-iterations}
\resizebox{\columnwidth}{!}{%
\begin{tabular}{|l|l|l|}
\hline
      & Iterations existing state-of-the-art \cite{dwivedi2023security}  & VISTA-CRYPT Iterations \\ \hline
%count & 46       & 40          \\ \hline
Mean  & 18445.41 & 9821.98    \\ \hline
Standard Deviation   & 17823.79 & 9821.9    \\ \hline
Min   & 85       & 17          \\ \hline
25\%   & 2522.5   & 1463        \\ \hline
Median   & 15474    & 8567        \\ \hline
75\%   & 28996.75 & 14801    \\ \hline
Max   & 65427    & 29928       \\ \hline
\end{tabular}%
}
\end{table}

\begin{figure}[t]
  \centering
  \begin{subfigure}{.475\columnwidth}
    \centering
    \includegraphics[width=\linewidth]{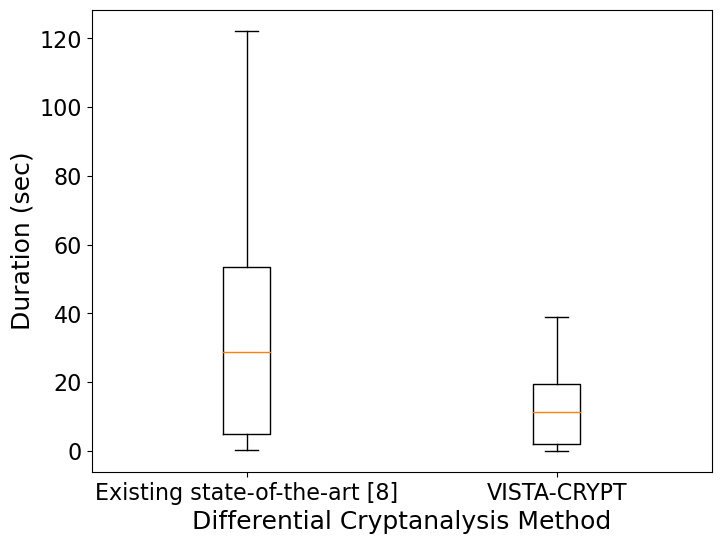}
    \caption{Duration performance comparison for SIMON $32$.}
    \label{fig:boxplot_duration}
  \end{subfigure}%
  \hfill
  \begin{subfigure}{.475\columnwidth}
    \centering
    \includegraphics[width=\linewidth]{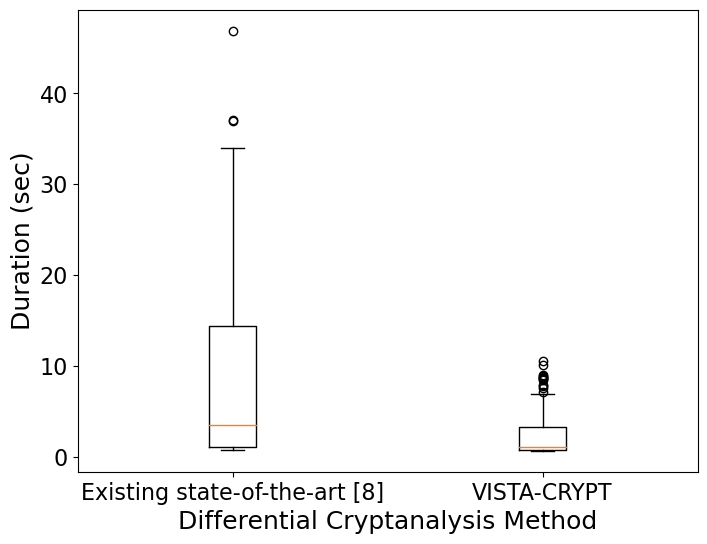}
    \caption{Duration performance comparison for SIMECK $32$.}
    \label{fig:simeck_boxplot_duration}
  \end{subfigure}%

  \caption{Performance comparison of differential cryptanalysis of the SIMON $32$ cypher \ref{fig:boxplot_duration} and SIMECK $32$ cypher \ref{fig:simeck_boxplot_duration} illustrating comparison between the existing state-of-the-art \cite{dwivedi2023security} and VISTA-CRYPT }
\end{figure}

\begin{figure}[t]
  \centering
  \begin{subfigure}{.475\columnwidth}
    \centering
    \includegraphics[width=\linewidth]{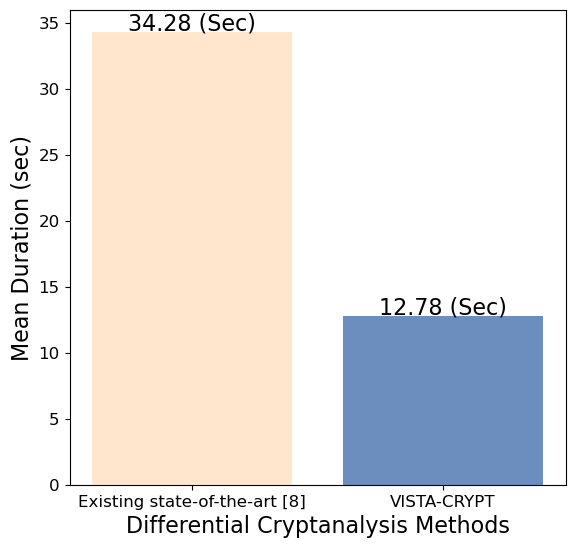}
    \caption{SIMON $32$ mean performance comparison}
    \label{fig:mean_comparison_duration}
  \end{subfigure}%
  \hfill
  \begin{subfigure}{.475\columnwidth}
    \centering
    \includegraphics[width=\linewidth]{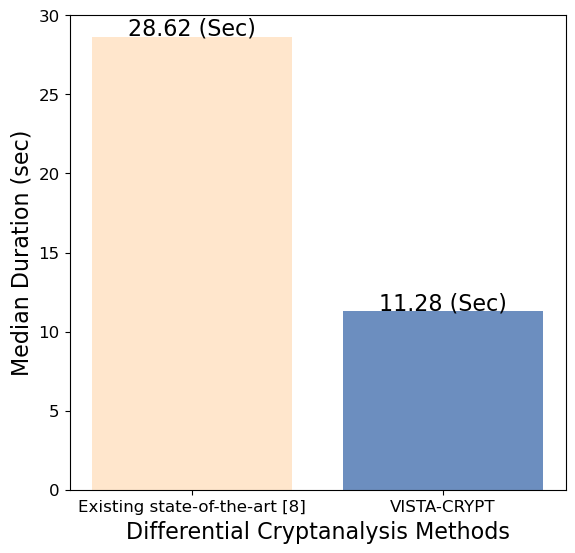}
    \caption{SIMON $32$ median performance comparison}
    \label{fig:median_comparison_duration}
  \end{subfigure}%

  \caption{Mean and median performance comparison of differential cryptanalysis of the SIMON $32$ cypher illustrating comparison between the existing state-of-the-art \cite{dwivedi2023security} (peach) and VISTA-CRYPT (blue)}
\end{figure}

Further investigation of the results highlights the efficiency gains of VISTA-CRYPT. As illustrated in Figures \ref{fig:boxplot_duration}, the spread of data has been reduced when compared to the technique developed by \cite{dwivedi2023security}, not only indicating variance reduction but also a significantly lower median duration. These results can be further visualised in Figures \ref{fig:mean_comparison_duration} and \ref{fig:median_comparison_duration} highlighting the efficiency gains. When performing a T-statistic test on the mean of both groups of experiments, we can determine that it returns $6.43$, indicating a significant difference between the original method and our proposed enhancements. This is further supported by a P-Value test, which is the probability of observing the given t-statistic if the null hypothesis is true. In our analysis, the P-Value returned a value of $1.17 \times 10^{-9}$, which is less than the commonly used $0.05$ significance level, allowing us to reject the null hypothesis and conclude that there is a significant difference between the two algorithm run times.

\begin{figure}[t]
  \centering
  \begin{subfigure}{.475\columnwidth}
    \centering
    \includegraphics[width=\linewidth]{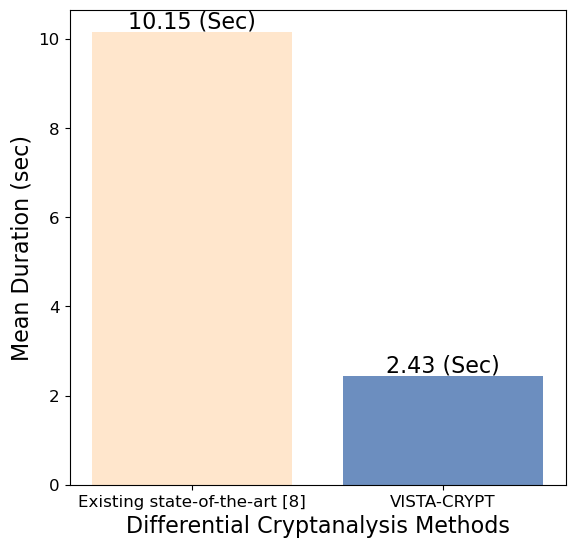}
    \caption{SIMECK $32$ mean duration performance comparison.}
    \label{fig:simeck_mean_comparison_duration}
  \end{subfigure}%
  \hfill
  \begin{subfigure}{.475\columnwidth}
    \centering
    \includegraphics[width=\linewidth]{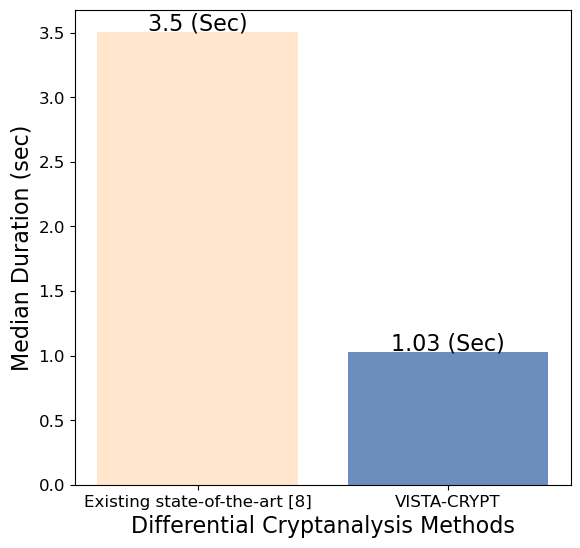}
    \caption{SIMECK $32$ median duration performance comparison.}
    \label{fig:simeck_median_comparison_duration}
  \end{subfigure}%

  \caption{Mean and median performance comparison of differential cryptanalysis of the SIMECK $32$ cypher illustrating comparison between the existing state-of-the-art \cite{dwivedi2023security} (peach) and VISTA-CRYPT (blue)}
\end{figure}

While the results of VISTA-CRYPT on the SIMON $32$ cypher appear conclusive, an analysis of the SIMECK $32$ cypher, although generally improved, is not as pronounced as those of the SIMON $32$ cypher with the identical target parameters. When applying the same target hamming weight as the SIMON $32$ cypher the cryptanalysis algorithms complete within several seconds on most experiments. This is despite the fact that one additional round of the cypher is being targeted, creating a more complex system to attack. Although the targeting of an additional round with a lower hamming weight results in a more complex cryptanalysis procedure, it potentially indicates a lower avalanche effect \cite{feistel1973cryptography}, replicating the findings of \cite{encarnacion2020modified} which compares to the results of \cite{yustiarini2022comparative}. Despite a lower level of improvement, the application of VISTA-CRYPT resulted in significant variance, path iteration and mean reductions, with only a marginal increase in the median time, which can be attributed to the sampling process. However, by reducing the target hamming weight the benefits of VISTA-CRYPT become apparent, demonstrating significant reductions to the mean and median of both time and number of iterations. Following the sanitisation of the data, with a target hamming weight of $28$, VISTA-CRYPT demonstrated an $76.06$ \% reduction in the mean time and an $63.97$ \% reduction in the mean number of iterations required to reach the target weight. As illustrated in Figures \ref{fig:simeck_mean_comparison_duration} and \ref{fig:simeck_median_comparison_duration}, the application of quota sampling on a lower target hamming weight corresponds with a significant time reduction. The reductions are further highlighted in Figure \ref{fig:simeck_boxplot_duration} which illustrates a significantly smaller spread of data, indicating a reduction in data variance between experiments. A summary of the results of the cryptanalysis of SIMECK $32$ is presented in Table \ref{tab:simeck32_analysis_28_weight}.

\begin{table}
\centering
\small
\caption{SIMECK 32 differential cryptanalysis: Comparison of the existing state-of-the-art \cite{dwivedi2023security} and VISTA-CRYPT Algorithm}
\label{tab:simeck32_analysis_28_weight}
\resizebox{\textwidth}{!}{%
\begin{tabular}{|l|l|l|l|l|}
\hline
      & Duration existing state-of-the-art \cite{dwivedi2023security} & Duration VISTA-CRYPT & Iterations existing state-of-the-art \cite{dwivedi2023security} & Iterations VISTA-CRYPT \\ \hline
%Count & 40                                     & 34                   & 40                                       & 34                     \\ \hline
Mean  & 10.15                               & 2.43             & 2970.4                              & 1070.1             \\ \hline
Standard Deviation   & 12.82                               & 2.49             & 3943.6                              & 1505.61             \\ \hline
Minimum   & 0.74                               & 0.64             & 11                                       & 3                      \\ \hline
25\%   & 1.063                               & 0.74             & 117.5                                    & 37                   \\ \hline
Median   & 3.5                               & 1.03             & 909                                      & 212                   \\ \hline
75\%   & 14.4                               & 3.23             & 4186.5                                   & 1565.75                    \\ \hline
maximum   & 46.9                               & 10.50             & 12824                                    & 5797                   \\ \hline
\end{tabular}%
}
\end{table}

\begin{figure}[t]
  \centering
  \begin{subfigure}{.475\columnwidth}
    \centering
    \includegraphics[width=\linewidth]{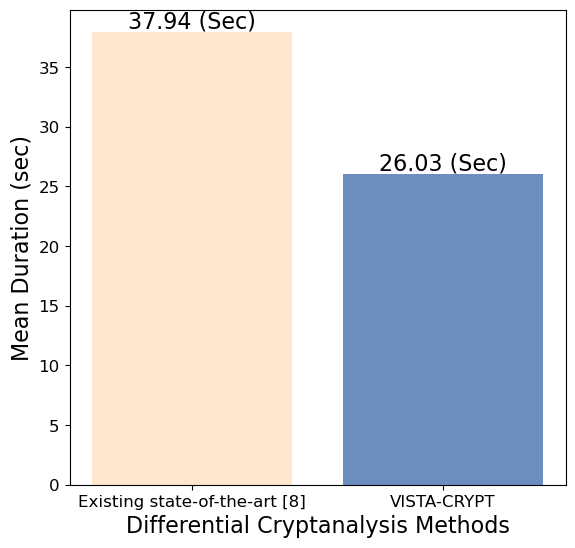}
    \caption{Mean two-way duration performance comparison for SIMON $32$.}
    \label{fig:simon_two_mean_comparison_duration}
  \end{subfigure}%
  \hfill
  \begin{subfigure}{.475\columnwidth}
    \centering
    \includegraphics[width=\linewidth]{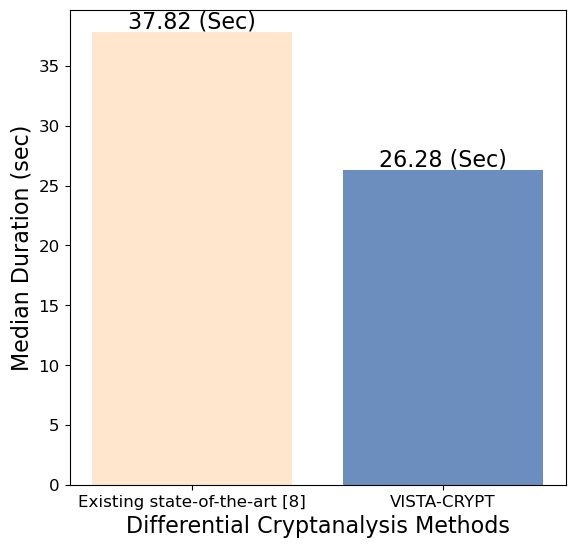}
    \caption{Median two-way duration performance comparison for SIMON $32$.}
    \label{fig:simon_two_median_comparison_duration}
  \end{subfigure}%

  \caption{Two-way SIMON $32$ differential cryptanalysis with target weight of $20$: Mean \ref{fig:simon_two_mean_comparison_duration} and median \ref{fig:simon_two_median_comparison_duration} performance comparison to reach goal hamming weight between the state-of-the-art \cite{dwivedi2023security} and VISTA-CRYPT }
\end{figure}

As highlighted by \cite{dwivedi2023security}, when the algorithm is split into two and run in a forward and backwards direction from the middle, additional efficiencies are gained. An analysis of the code provided in their GitHub further validates their results. When assessing the SIMON cryptanalysis two-way file, several observations that support additional efficiencies are apparent. First, the cryptanalysis has been defined to attack eleven rounds of the cypher, rather than ten rounds as in the one-direction algorithm. The number of rounds to attack is defined as six in the reverse direction and five in the forward direction. Secondly, the target hamming weight has been defined as twenty, which is considerably lower than the target weight of $32$ defined in the one-direction algorithms. To assess the performance of VISTA-CRYPT on the two-direction algorithm, it was necessary to test the performance with the variables as defined by \cite{dwivedi2023security} and analyse the comparisons between the existing state-of-the-art technique and VISTA-CRYPT. The results of our experimental analysis further support the findings of improved efficiency outlined above. As illustrated in Figure \ref{fig:simon_two_mean_comparison_duration}, the application of VISTA-CRYPT reduces the mean duration of differential cryptanalysis of the SIMON $32$ cypher when the target hamming weight has been defined at $20$, which is the goal weight defined in \cite{dwivedi2023security}'s GitHub code. The results are additionally duplicated with the median duration time, as shown in Figure \ref{fig:simon_two_median_comparison_duration}. Although the number of iterations is similar between both techniques, it is clearly evident that VISTA-CRYPT demonstrates significant performance improvements over the existing state-of-the-art technique.

\begin{table}
\centering
\caption{Weight attained over time comparing the existing state-of-the-art and VISTA-CRYPT}
\label{tab:simon-two-way-round-compare}
\begin{tabular}{|l|p{3.2cm}|p{3.2cm}|p{3.2cm}|}
\hline
Weight \newline & Duration Existing \newline state-of-the-art (sec) & Duration \newline VISTA-CRYPT  (sec) \\ \hline
44           & N/A                                & 0.0009899            \\ \hline
36           & N/A                                & 0.0019972            \\ \hline
32           & 0.0039995                          & 0.0049953            \\ \hline
30           & 0.0571863                          & N/A                  \\ \hline
28           & 0.1302797                          & 3.3259256            \\ \hline
26           & 0.3343815                          & 3.4711129            \\ \hline
24           & 0.5183801                          & 5.8765139            \\ \hline
22           & 0.5673296                          & 5.8785143            \\ \hline
20           & 37.9628699                         & 19.0578536           \\ \hline
\end{tabular}
\end{table}

Although performance enhancements are evident from the results of our analysis, an investigation of the time taken to reach specific weights reveals interesting insights into the cryptanalysis progression. As shown in Table \ref{tab:simon-two-way-round-compare}, although VISTA-CRYPT attains the goal weight significantly faster than the existing state-of-the-art, it is initially inherently slower with higher weight than the existing method, illustrating poorer initial performance. Indeed, as illustrated in Table \ref{tab:simon-two-way-round-compare}, weights above the target are attained later than the existing work by the authors of \cite{dwivedi2023security}, however, as the weight reduces, the performance of VISTA-CRYPT exceeds that of the existing state-of-the-art. This characteristic is repeatable in every experiment conducted. Of note, however, is that when the target weight is reduced below $20$ the performance of both implementations is degraded significantly, with the existing state-of-the-art failing to produce results and VISTA-CRYPT exceeding the sample size without conclusively reaching the desired goal weight.

\begin{figure}[t]
  \centering
  \begin{subfigure}{.475\textwidth}
    \centering
    \includegraphics[width=\linewidth]{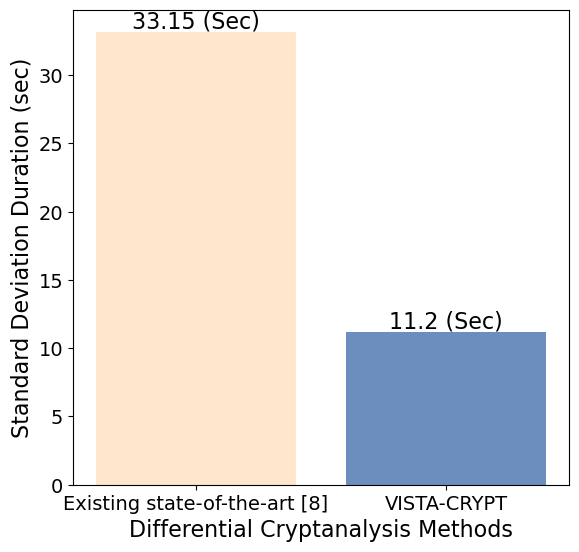}
    \caption{Duration standard deviation for SIMON $32$.}
    \label{fig:simon_32_duration_sd}
  \end{subfigure}%
  \hfill
  \begin{subfigure}{.475\textwidth}
    \centering
    \includegraphics[width=\linewidth]{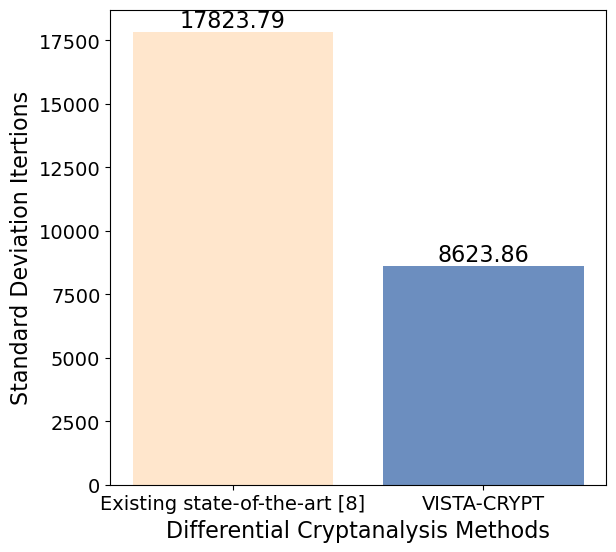}
    \caption{Iterations standard deviation for SIMON $32$.}
    \label{fig:simon_32_iteration_sd}
  \end{subfigure}%

  \caption{Standard deviation of the duration \ref{fig:simon_32_duration_sd} and number of iterations \ref{fig:simon_32_iteration_sd} for SIMON $32$ differential cryptanalysis comparing the existing state-of-the-art \cite{dwivedi2023security} and VISTA-CRYPT }
\end{figure}

Further analysis of the performance enhancements implemented using VISTA-CRYPT highlights additional improvements beyond savings of time and the number of iterations. While our analysis has illustrated a reduction in variance as well as significant reductions in both time and the number of iterations, a study of the effects of our technique on the standard deviation of the experiments illustrates improvements in the quality of the results. As shown in Figures \ref{fig:simon_32_duration_sd} and \ref{fig:simon_32_iteration_sd}, the standard deviation, which is a measure of how far the data varies from the mean \cite{wan2014estimating}, shows a substantial reduction following the implementation of our technique. This demonstrates that prior to VISTA-CRYPT, the outcome of each experiment had a larger range of results, highlighting a distinct shortcoming of heuristic search methods. With the application of our techniques, experiments demonstrate more consistent results, allowing for further investigation into the discovery of potential characteristics of the search that may assist in additional round exploitation, or further improvements in efficiency.

The performance enhancements of VISTA-CRYPT coupled with the analysis of the GitHub code provided by \cite{dwivedi2023security} raise thought-provoking questions about further modifications of the number of rounds to attack and target hamming weight. To assess the efficacy and efficiency of VISTA-CRYPT further, we modified the number of rounds to attack to $16$, which is one additional round than \cite{dwivedi2023security} successfully attacked, with a goal hamming weight of $36$. To investigate the adjustments, the number of experiments was reduced to ten for each technique, as opposed to $193$ for the above analysis undertaken in Section \ref{Sec:Methods}. A decision was made to examine only ten experiments due to the prolonged execution time required to undertake differential cryptanalysis of additional rounds. Although this goal weight is higher than the weight defined by the size of the block cypher, we were able to successfully execute the attack on all $16$ rounds with a mean time reduction of $19.55$ \% over the existing technique. However, while the experiments modifying the number of rounds and target weight proved to be successful, they exhibited inefficiencies that made lowering the goal weight to the size of the cypher infeasible.

Although the existing implementation of the NMCS on the lightweight cyphers devised by \cite{dwivedi2023security} is comparatively efficient, an analysis of our proposed sampling methods demonstrates considerable performance improvements. Our approach of reducing variance and population size has resulted in overall improvements in time efficiency without a degradation to the overall results. A thorough analysis of the results shows a mean time reduction of $62.72$ per cent when using quota sampling, extending to $76.06$ per cent in SIMECK with a target hamming weight of $28$ in the one-direction algorithm. However, the two-way algorithm demonstrated a lesser impact than the one-way method with a mean reduction in time at $31.39$ per cent. Although not as significant a time reduction, it remains substantially large enough to be considered a modification worth implementing. Statistical analysis with the t-test and P-Value further supports the results presented by our strategy. The results were further enhanced with an analysis of the two-direction cryptanalysis, demonstrating similar performance improvements.

\section{Preliminary Graph based analysis }\label{Sec:GraphAnalysis}

\begin{figure*}[t]
     \centering
     \begin{subfigure}[b]{.475\textwidth}
         \centering
         \includegraphics[width=0.8\columnwidth]{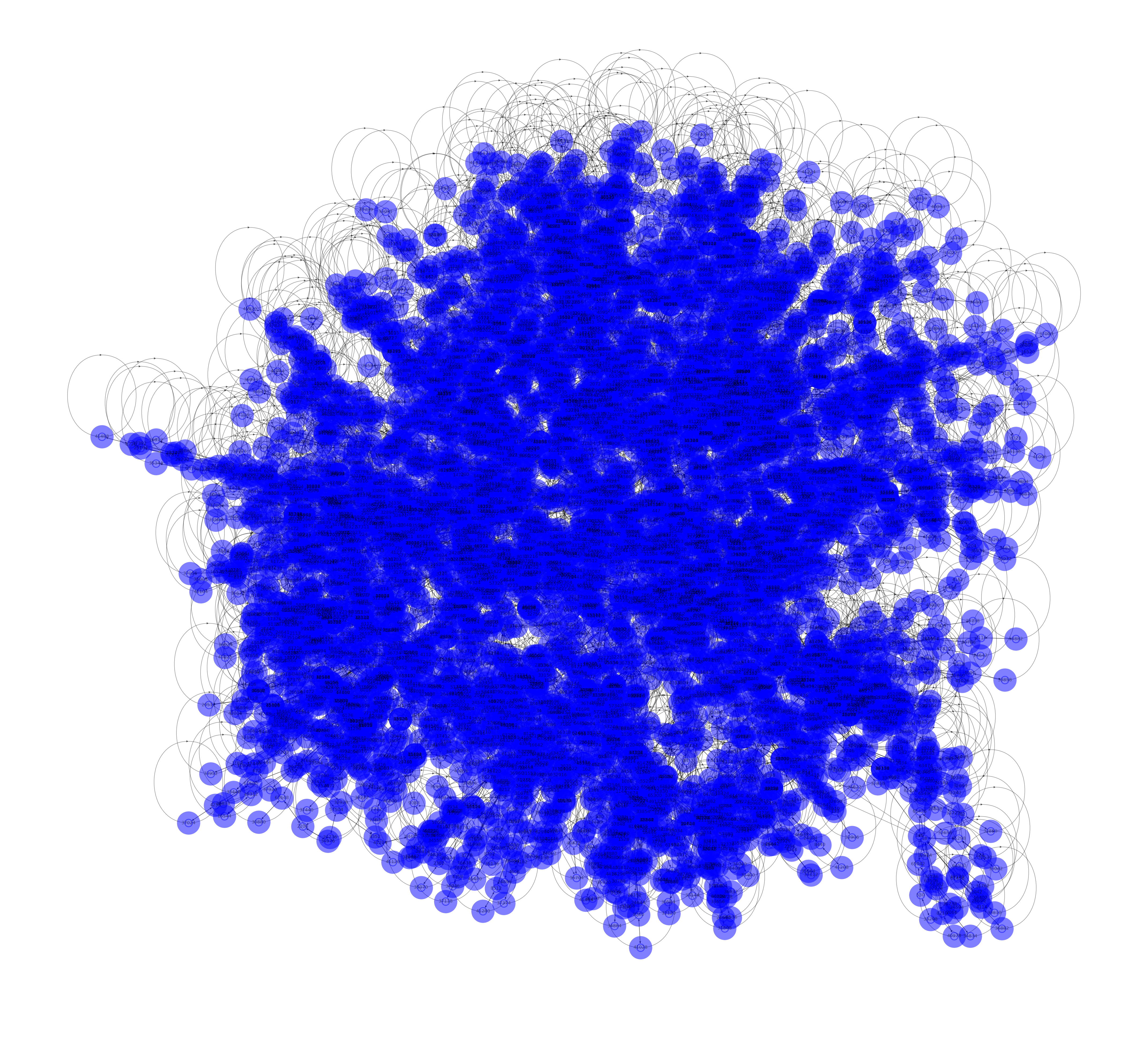}
         \caption{Knowledge graph of the full distribution of output differentials for SIMON $32$.}
         \label{fig:graph_full_dist}
     \end{subfigure}%
     \hfill
     \begin{subfigure}[b]{.475\textwidth}
         \centering
         \includegraphics[width=0.8\columnwidth]{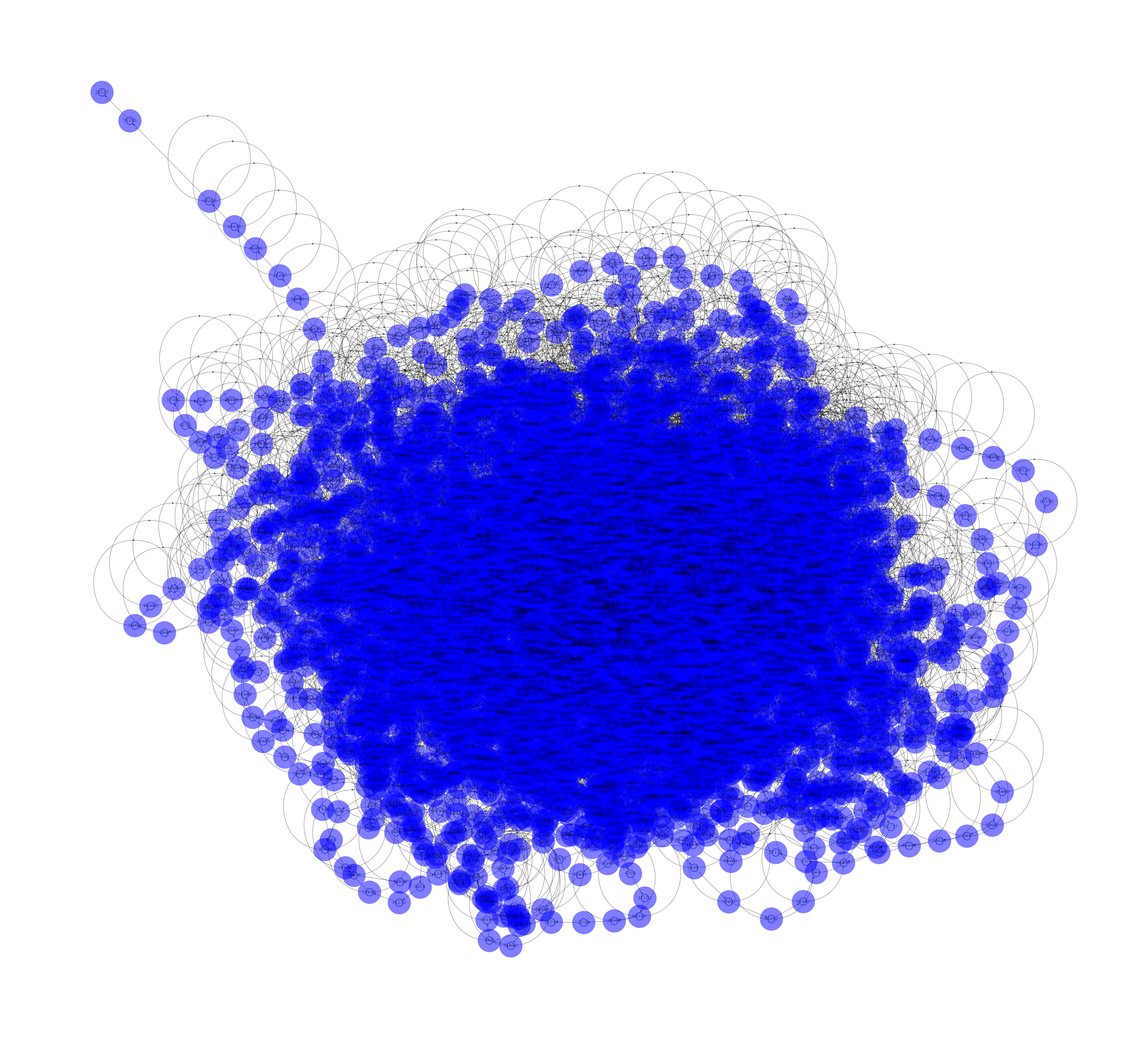}
         \caption{Knowledge graph of the sample distribution of output differentials for SIMON $32$.}
         \label{fig:graph_sample_dist}
     \end{subfigure}
        \caption{Knowledge graph illustrating relationships between output differentials of the existing state-of-the-art \ref{fig:graph_full_dist} and VISTA-CRYPT \ref{fig:graph_sample_dist}. }
        \label{fig:graph_sample_orig}
\end{figure*}

Analysing data using graph databases offers a powerful approach to uncovering intricate relationships and patterns within complex datasets. As we have demonstrated through our sampling method, although the supplied data represents a list of output differentials of a cryptanalysis function, the data within the lists can be utilised as a dataset, and as such, can be explored for an understanding of data relationships and optimisation analysis. Graph databases, such as Neo4j \cite{Neo4j2020}, enable the representation of data as nodes and edges, which facilitates the modelling of entities and their connections. The graph-based approach allows for the exploration of connections between entities and the traversal of relationships, providing valuable insights into the structure and dependencies within the data. As differential cryptanalysis is the study of how changes to an input can propagate changes to an output, the relationship between these changes can be applied to a graph database. The application of graph databases to build knowledge graphs could be applied to differential and linear cryptanalysis. As noted in \cite{xia2021graph}, analyzing complex and large-scale structures generated through cryptanalysis can enable graph learning to capture intricate relationships among vertices.

The creation of a knowledge graph for a preliminary analysis requires the selection of an appropriate tool to conduct a study of the complex nature of cryptography and cryptanalysis and the relationship between differentials. According to \cite{Neo4j2021What}, Neo4j offers an efficient and streamlined approach to identifying connections between data points. This is due to the connections not being processed at query time, as with other programming languages such as Python, and instead, the connections between data points are stored directly in the graph database. When compared to traditional relational databases, such as MySQL, Neo4j can produce results more than $1,135$ times faster than MySQL \cite{Neo4j2012how}.   While Neo4j is the most widely used software solution for graph databases \cite{li2023cybersecurity}, the cloud infrastructure and node limitations of the software can inhibit the study of vast and complex structures derived from differentials. For this reason, we have deployed the development of a knowledge graph using the Python library NetworkX which allows for the creation of knowledge graphs within Python \cite{hagberg2008exploring}.

For this preliminary analysis, we have opted to investigate the output differentials, comparing the structures of the full distribution of differentials and our sampling method. Performing a graph-based analysis of the output differentials of the existing state-of-the-art method as shown in Figure \ref{fig:graph_full_dist} and our sampling method illustrated in Figure \ref{fig:graph_sample_dist}, allows for the analysis of the structures of both models. Although both graphs demonstrate an extremely tight clustering of differentials, it is still possible to infer characteristics and differences between both models, drawing conclusions based on the centrality, modularity density and connectivity of the respective graphs as well as potential future research opportunities. The analysis can further establish a connection between the performance differences between both techniques of differential cryptanalysis. 

Exploring the graph shown in Figure \ref{fig:graph_full_dist} illustrates that the network contains a strong central core due to the concentration of nodes and connectivity. However, also visible in the graph are small vacant pockets which suggests there are subgroups within the network that are relatively isolated or have specific functions not closely integrated within the main network flow. The peripheral branches illustrate relative density but with less centrality to the core nodes. The peripheral branches could represent components that are less critical to the overall function of cryptanalysis. When inferring performance possibilities from the graph, the density of the peripheral branches points towards lower levels of failure from the branches as the core remains intact, however, isolated pockets may be points of inefficiency if they are critical the the differential process. 

In contrast, the graph shown in Figure \ref{fig:graph_sample_dist} shows an extremely dense and highly connected central core where nodes are highly dependent on each other. The peripheral branches are less densely populated and more dispersed, representing areas of the network that are not as specialised and used as frequently as the inner core. Of note, the most distinct peripheral branch extends further from centrality than other peripheral branches illustrating a highly specialised area of the network seldom utilised and largely uncritical to the main function of cryptanalysis. The peripheral branches with circular connections that appear closer to centrality indicate secondary processes that support the core but are not as critical to the network operation. When evaluating the performance of this network, the densely populated central core should result in highly efficient outcomes, however, it may be vulnerable to a cascading failure in the event that a central node fails due to the high level of interdependence within the core.

The key differences between the two knowledge graphs are summarised as follows. The graph illustrated in Figure \ref{fig:graph_sample_dist} suggests a high degree of connectivity between nodes due to the highly concentrated inner core that will result in quicker dissemination of information and resources through the network. In essence, it provides a more efficient and faster progression through the differential due to the density of connected nodes. However, while it is highly centralised and more efficient, it is vulnerable to a single point of failure that can inhibit overall outcomes. Comparatively speaking, the graph illustrated in Figure \ref{fig:graph_full_dist} is more modular with isolated pockets, illustrating that while it is slower and less efficient, it will be less vulnerable to localised failures than the graph shown in Figure \ref{fig:graph_sample_dist}. Although the graph illustrated in Figure \ref{fig:graph_sample_dist} is more prone to failure than the graph illustrated in Figure \ref{fig:graph_full_dist}, the highly congested inner core combined with the experimental results illustrated in Section \ref{Sec:Results} demonstrates that the chosen method of sampling differentials is beneficial for differential cryptanalysis.

\section{Discussion }\label{Sec:Discussion}

The results we present show that the addition of quota sampling in NMCS significantly improves the efficiency of the search algorithm, echoing previous studies utilising a derivative of this technique. However, despite the promising results of VISTA-CRYPT, the technique still relies on simple random sampling which has several disadvantages. Although the population has been reduced, the process of randomly selecting a path from the sample can lead to inconsistent results, requiring many experiments to deduce the mean and median performance metrics. However, opportunities in future research may be possible through investigating pseudorandomness \cite{vadhan2012pseudorandomness} within the heuristic and if adjustments to the seed value through machine learning \cite{kelsey1998cryptanalytic} can produce consistent results with improved efficiency. While uncommon and considered an outlier event, inefficient paths can occur due to the random nature of the heuristic search. This is further highlighted by the preliminary graph-based analysis demonstrating a high density of differentials around the core that may produce failures, resulting in inefficient explorations. Should a random event land on one of these differentials, efficiency gains may be lost. Nevertheless, the process of random sampling reduces the total number of paths to explore when compared to an exhaustive search of all paths, resulting in an overall significantly more efficient endeavour. When coupled with quota sampling techniques, as demonstrated with VISTA-CRYPT, the challenges presented by random sampling can be significantly reduced, providing state-of-the-art results in differential cryptanalysis.

In developing a streamlined and efficient technique of differential cryptanalysis, we have made the following five contributions. First, we identified the limitations of random sampling in MCS when applied to differential cryptanalysis undertaken by the existing state-of-the-art technique. Having identified this limitation, our second contribution introduced proportional representative sampling of the output differentials that resulted in a reduction in the variance within the differentials. This in turn improves algorithm efficiency with fewer iterations required to reach the desired hamming weight which is reflected in time reductions. Through a comparative analysis of the data, our third contribution demonstrated significant reductions in execution time and the number of iterations, with savings in time efficiency of up to $76$ \% and a reduction of $63.97$ \% in the number of iterations, addressing RQ2. Our fourth contribution demonstrates that the early termination of inefficient experiments allows for the faster production of synthetic datasets for use in quantitative analysis. By discontinuing experiments that are taking longer to process with data that may be removed during data cleaning, the process of conducting many experiments is accelerated. Additionally, in a real-world and real-time environment with an attack conducted against low-powered IoT devices whose resources are constrained, the discontinuation of an inefficient attack could result in a successful outcome at a later time without compromising the quality of service of the target device. Finally, the fifth contribution is a preliminary graph-based analysis identifying that while the sampling method produces a more dense and closely connected graph illustrating an improvement in speed and efficiency, there is an increased potential for a cascading failure if a central node fails. With this observation in mind, future research opportunities exist in identifying and isolating clusters of nodes that can lead to failure, potentially further improving performance and efficiency.

\section{Conclusion and future research }\label{Sec:Conclusion}

In this paper, we have identified limitations of simple random selection with the existing state-of-the-art technique of Nested Monte-Carlo Search (NMCS) in cryptanalysis that hinders algorithm efficiency. Through thorough experimentation, we have identified sampling techniques suitable for improving the efficiency of NMCS in a differential cryptanalysis setting. By ensuring a proportional representation of the output differentials is present in the sample, we have demonstrated that the size of the output differential population is significantly smaller than the full distribution and that the variance within the list of differentials is also reduced, leading to efficiency gains.  Although a limited number of experiments exhibit performance worse than the existing implementation, the early termination of experiments that encounter a suboptimal path results in overall performance gains, ensuring quality of service in a real-world environment is maintained. Through an extensive and detailed analysis of the existing state-of-the-art technique and our proposed enhancements, we have demonstrated significant reductions in time of up to $76$ \% with quantifiable reductions in the number of iterations for the one-way algorithm. We additionally established that as the target hamming weight is reduced the performance improves over the existing state-of-the-art technique, as illustrated with the SIMECK $32$ cypher. Further, through our initial graph-based analysis we have identified both strengths and weaknesses of the proposed technique and identified potential areas of future research which may produce more efficient results. We have demonstrated that the algorithm for stratifying differentials is simple and effective, allowing for its potential application against other block cyphers.

 \nocite{*} 
\bibliographystyle{unsrt}
\bibliography{references}  %%% Uncomment this line and comment out the ``thebibliography'' section below to use the external .bib file (using bibtex) .

%%% Uncomment this section and comment out the \bibliography{references} line above to use inline references.
% \begin{thebibliography}{1}

% 	\bibitem{kour2014real}
% 	George Kour and Raid Saabne.
% 	\newblock Real-time segmentation of on-line handwritten arabic script.
% 	\newblock In {\em Frontiers in Handwriting Recognition (ICFHR), 2014 14th
% 			International Conference on}, pages 417--422. IEEE, 2014.

% 	\bibitem{kour2014fast}
% 	George Kour and Raid Saabne.
% 	\newblock Fast classification of handwritten on-line arabic characters.
% 	\newblock In {\em Soft Computing and Pattern Recognition (SoCPaR), 2014 6th
% 			International Conference of}, pages 312--318. IEEE, 2014.

% 	\bibitem{hadash2018estimate}
% 	Guy Hadash, Einat Kermany, Boaz Carmeli, Ofer Lavi, George Kour, and Alon
% 	Jacovi.
% 	\newblock Estimate and replace: A novel approach to integrating deep neural
% 	networks with existing applications.
% 	\newblock {\em arXiv preprint arXiv:1804.09028}, 2018.

% \end{thebibliography}

\end{document}